\documentclass[conference]{IEEEtran}
\IEEEoverridecommandlockouts
\usepackage{cite}
\usepackage{amsmath,amssymb,amsfonts}
\usepackage{algorithmic}
\usepackage{graphicx}
\usepackage{textcomp}
\usepackage{subfigure}
\usepackage{xcolor}
\usepackage{makecell}
\usepackage{mathptmx} 
\usepackage{tabularx}
\usepackage {booktabs} 
\mathchardef\mhyphen="2D
\def\BibTeX{{\rm B\kern-.05em{\sc i\kern-.025em b}\kern-.08em
    T\kern-.1667em\lower.7ex\hbox{E}\kern-.125emX}}
\begin{document}

\title{Twotier - A Layered Analysis of Backbone Members in a Moderate Sized Community Sports Organization\\
\thanks{Qingran Wang and Jia Yu contributed equally to this paper. We would like to thank all members of the SJTU Health community for their selfless commitments in building a strong community.}
}
\author{Qingran~Wang, Jia~Yu, Mengjun~Ding,  Weiqiang~Sun, ~\IEEEmembership{Senior~Member,~IEEE}
}
\maketitle

\begin{abstract}
Backbone members are recognized as essential parts of an organization, yet their role and mechanisms of functioning in networks are not fully understood. In this paper, we propose a new framework called Twotier to analyze the evolution of community sports organizations (CSOs) and the role of backbone members. Tier-one establishes a dynamic user interaction network based on grouping relationships, and weighted k-shell decomposition is used to select backbone members. We perform community detection and capture the evolution of two separate sub-networks: one formed by backbone members and the other formed by other members. In Tier-two, the sub-networks are abstracted, revealing a core-periphery structure in the organization where backbone members serve as bridges connecting all parts of the network. Our findings suggest that relying on backbone members can keep newcomers actively involved in rewarding activities, while non-rewarding activities solidify relations between backbone members.
	
\end{abstract}


\begin{figure*}[t]
	\centering
	\includegraphics[width=7.19in]{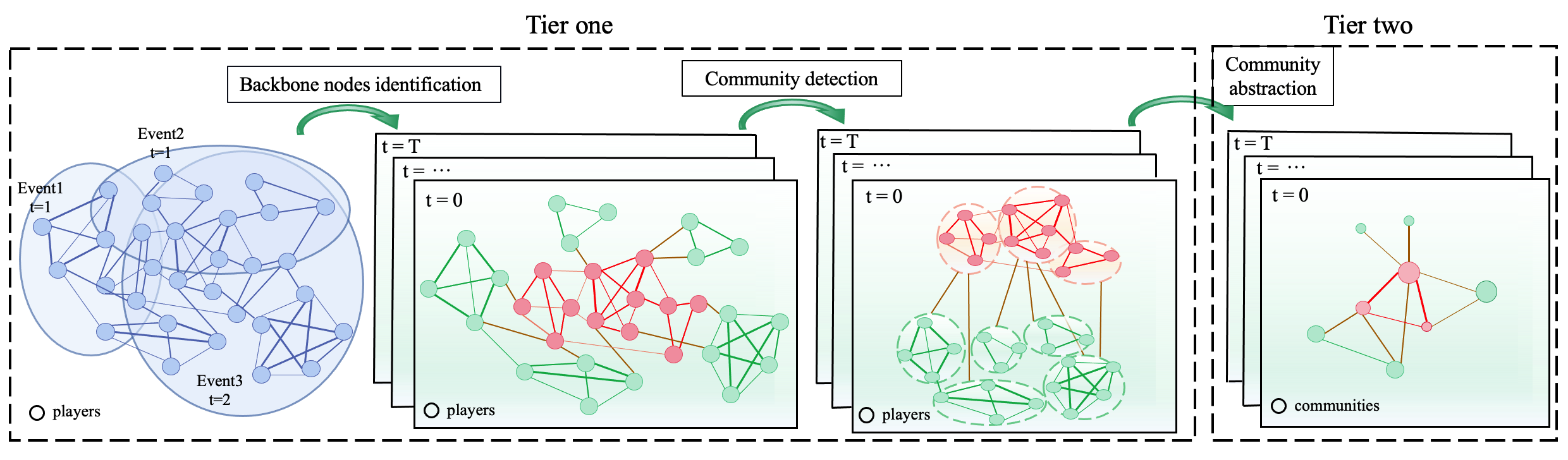}
	\caption{The Twotier Framework}
	\label{fig_framework}
\end{figure*}

\section{Introduction}
Community sports organizations (CSOs) are non-profit and voluntary organizations whose primary responsibility are to provide sports services to their members, often with low threshold to entry \cite{misener2014support,van2020community}.
Despite the huge physical and psychological benefits CSOs can bring to their community members, the development of CSOs are often constrained by their voluntary nature and the limited resource available to them \cite{westberg2022promoting, doherty2019organizational}. It is thus important to understand the development principle of CSOs such that the limited resource may be put to the most effective use. 

It has always been intuitively felt that there is usually a group of highly active and influential people who actuate and drive the development of a network. In the process of product marketing based on the human interaction network, marketers will take the nodes occupying the structural hole position in the network as the influential initial node, in order to achieve the greatest influence in the network  \cite{zhu2017new}. In online social networks such as Weibo, Twitter, etc., users with a large number of followers are considered as influential users, and the topics they publish tend to generate great network effects \cite{2020The, zhao2021propagation}. Similarly, in CSOs, there are also some influential users who have ``the right and the ability to influence in an indirect or intangible way" \cite{zhao2014finding}, and their presence and activities have significantly higher effects on the operation and development of the organization.  Backbone members are important nodes in a network that are well-connected to other members and play a crucial role in facilitating communication and information flow. They are defined as members who are relatively more important, active, and have a greater number of friends compared to other members. The identification and analysis of backbone members can provide insights into the structure and dynamics of the network, which is valuable for understanding its behavior and performance.

The problem of vital node identification has attracted increasing attention in different fields \cite{lu2016vital}. 
Typically, researchers build user social networks based on participant interaction data collected over a period of time and then work to identify key nodes in the network. In this scenario, various centrality measures \cite{de2019general, yu2022epidemic, santoro2022onbra} such as degree centrality, closeness centrality, and betweenness centrality can be used to indicate the importance of nodes. With the rich set of metrics introduced in \cite{das2018study,  rombach2014core, kitsak2010identification, garas2012k} we can also identify important nodes in CSOs. However, little attention has been paid to the role that backbone members play, nor the mechanisms by which they function in the network. At the same time, studies are often focused on the group of backbone members themselves, and interactions between the backbone and non-backbone members are largely neglected. In addition to the internal forces generated by the backbone members, external interventions such as rewards and penalties may also be crucial for network development \cite{zhou2020formation, yu2021community}. Targeting interventions on leaders have been shown to be more effective than applying them to random individuals for community health campaigns \cite{shafiq2013identifying}. Understanding the mechanisms by which internal forces work can help us better implement external interventions \cite{ott2018strategic}. And, if these two forces work together, they can bring even greater developmental benefits. 

In this research, with longitudinal data recorded, we focus on the development of CSOs, with a particular emphasis on backbone members, defined as the top X\% of influential members based on coreness centrality. We introduce Twotier, a new framework for analyzing dynamic networks, which allows us to study both the evolutionary characteristics of components and their connections. Our main finding is that backbone members play a critical role as the trunk of the network, while others act as leaves and are regularly updated. Rewarding activities and backbone members are essential for organization expansion, while non-rewarding activities solidify the backbone group.
The main contributions of this work are three-fold. Firstly, we introduce Twotier, a novel mathematical framework for analyzing dynamic networks. Secondly, we demonstrate its applicability in a moderate-sized CSO. Finally, using our framework and numerical results, we provide practitioners with tailored approaches to improve outcomes for different groups within their organization.

The remainder of this paper is organized as follows. In Section II, we introduce Twotier, the main method for network analysis in this work, and explain its specific procedure. In Section III, we present an overview of the dataset and the experimental results obtained in each tier, including the role of backbone members in network development and their performance under external factors. Section IV provides an overview of related work. Finally, in Section V, we conclude the paper.

\begin{table}[t]
	\caption{Symbols and Description}
	\begin{center}
		\begin{tabular}{c|c}
			\hline
			\textbf{Symbols} & \textbf{Description}                                                                                           \\ \hline
			$G_{t}$          & \begin{tabular}[c]{@{}c@{}}Snapshot network with players as nodes \\ in the $t$-th time frame\end{tabular}     \\
			$G^{com}_{t}$    & \begin{tabular}[c]{@{}c@{}}Snapshot network with communities as nodes \\ in the $t$-th time frame\end{tabular} \\
			$w^{ij}_{t}$     & The weight of an edge with node $i$ and $j$ in $G_{t}$                                                         \\
			$n_t^i$          & The set of neighbors of node $i$ in $G_{t}$                                                                 \\
			$d_{t}^{i}$      & The number of neighbors of node $i$ in $G_{t}$                                                                 \\
			$D_{t}^{i}$      & Weighted degree of node $i$ in $G_{t}$                                                                         \\
			$k_{t}^{i}$      & The layer of node $i$ in $G_{t}$                                                                               \\
			$I_{t}^{i}$      & Influence of node $i$ in $G_{t}$                                                                               \\
			$Q_{t}$          & Modularity of $G_{t}$                                                                                          \\ \hline
		\end{tabular}
	\end{center}
	\label{tab_symbols}
\end{table}

\section{Twotier: A Layered Analysis on CSOs}
This section introduces the Twotier framework, which analyzes the role of backbone members in a moderate-sized community sports organization. In Tier-one, we build a dynamic network based on team-wise links between members and classify them into two groups: backbone members and general members, using the dynamic W-KS algorithm to calculate their influence. Community detection is performed to transition from individuals to communities. In Tier-two, we analyze the evolutionary regularities of different types of communities by abstracting the dynamic network into the network of communities extracted in Tier-one. The framework is illustrated in Fig.~\ref{fig_framework}. To explore the influence of different types of activities on organizational development, we separate the network into two sub-networks: one formed under rewarding activities and the other under non-rewarding activities. Table \ref{tab_symbols} 
summarizes the symbols used in this paper.

\subsection{Tier-one Analysis}

\begin{table*}[t]
	\caption{Evolution Events}
	\begin{center}
		\begin{tabular*}{\textwidth}{c|c|c}
			\toprule[1pt]
			\makebox[0.15\textwidth] [c] {\textbf{Evolution events}}  &  \textbf{Properties} & \textbf{Attribute}\\
			\hline
			Form & \makecell[l]{All members of the group have not appeared in  previous time frames.} & $V$\\
			\hline
			Re-emerge & \makecell[l]{Members of the group are not present in the previous time frame $T-1$, but has appeared earlier.}  & $V$\\
			\hline				
			Continue & \makecell[l]{A group continues its existence, when two groups in the consecutive time frames are identical, \\or when two groups differ only by few nodes but their size remains the same.} & --\\
			\hline
			Grow & \makecell[l]{A group grows when some new nodes have joined the group, \\making its size bigger than in the previous time frame $T-1$.} & $S$\\
			\hline					
			Split & \makecell[l]{A group splits into two or more groups in the next time frame $T+1$, \\when some groups from time frame $T+1$ consist of members of one group from the previous time frame $T$.} & $S$\\
			\hline
			Merge & \makecell[l]{A group has been created by merging several other groups, \\when one group from time frame $T+1$ consists of two or more groups from the previous time frame $T$.} & $S$ \\
			\hline
			Shrink & \makecell[l]{A group shrinks when some nodes have left the group, \\making its size smaller than in the previous time frame $T-1$.} & $S$\\
			\hline
			\makecell[c]{Suspend} & \makecell[l]{A group  does not emerge in the next time frame $T+1$, but will re-appear later. } & $V$\\
			\hline
			Dissolve & \makecell[l]{A group ends its life and will not re-appear in future time frames.} & $V$\\
			\bottomrule[1pt]
		\end{tabular*}
		\label{tab_event}
	\end{center}
	\vspace{-2mm}
	\footnotesize{The properties of different types of evolution events are based on \cite{brodka2013ged}.\\$V-$ Events that bring significant structural changes to the network, mainly through adding/removing nodes from the network, \\$S-$ Events that cause marginal structural changes to the network, mainly through adding/removing links between existing nodes.}
\end{table*}

\subsubsection{Evaluating Social Influence by Dynamic W-KS}
In a CSO with teaming-based relationships,  user interactions can change over time, resulting in a dynamic network. Therefore, it is not advisable to apply vital node identification approaches designed for static networks. For example, it may be challenging to determine the importance of a node that is active during some time periods but inactive in others. To address this issue, we extend the weighted k-shell decomposition method to be applied on dynamic networks as a series of static networks.

Considering the duration of activities and the fact that the study is conducted under a six-year time span, we take a three-month time window to build a dynamic network containing 24 consecutive equal-length time frames, in each of which the network is considered non-evolving. The validity of this partitioning approach has been demonstrated in our previous work \cite{yu2021community}. In this case, the network is expressed as 
\begin{equation}
	G=\{G_{t}=(V_{t}, E_{t}), \ for\ all\ t\ in\ [0, T]\},
	\label{eq_Graph}
\end{equation}
where $V_{t}$ is the node set and $E_{t}$ is the set of edges. The weights of the edges are assigned with the number of links that connect the same $node\ pair$ within a time frame.

\begin{figure}[t]
	\centering
	\subfigure[]{\includegraphics[height=0.324\textwidth]{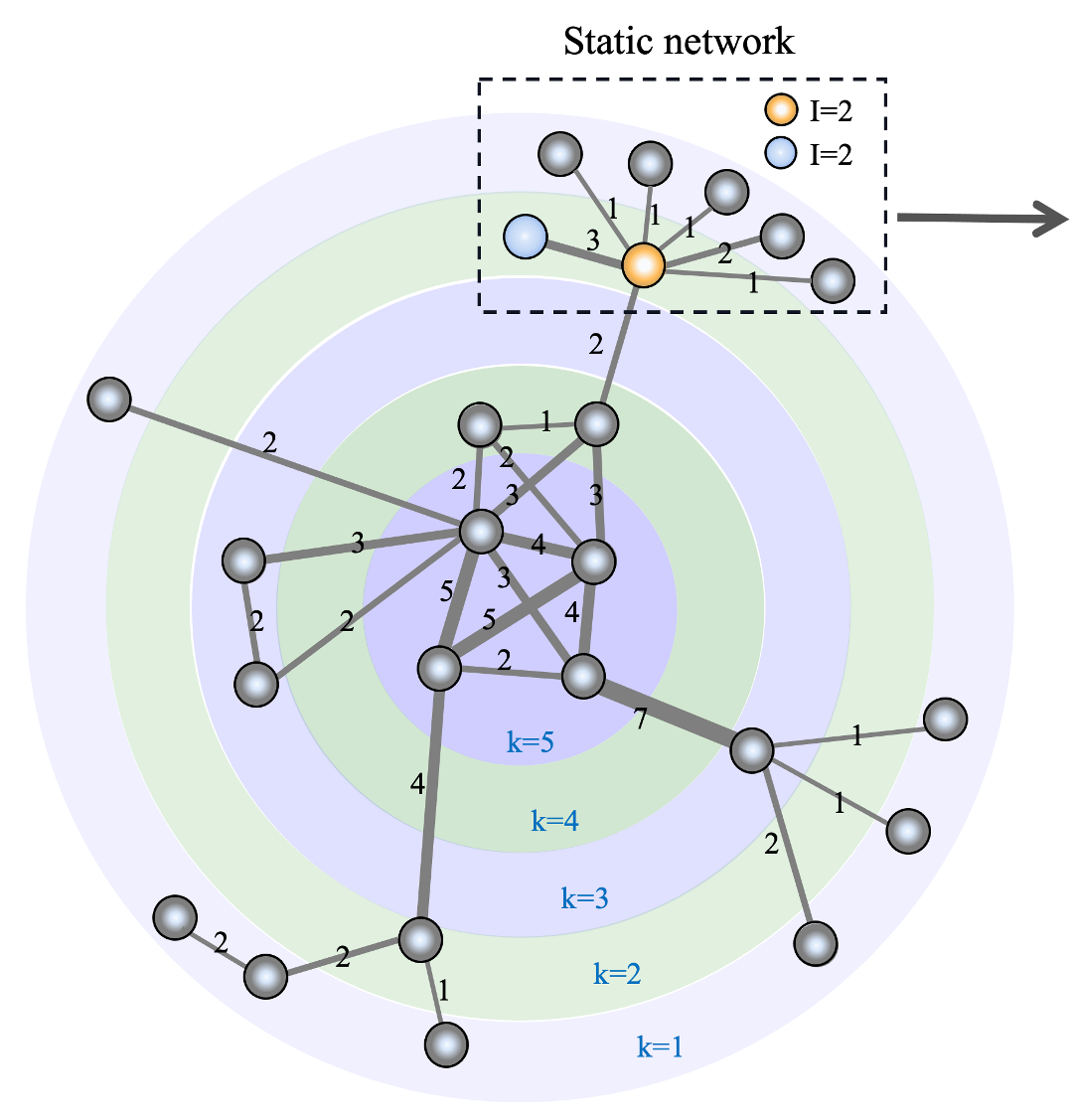}%
		\label{fig_kshell1}}
	\hfil
	\subfigure[]{\includegraphics[height=0.324\textwidth]{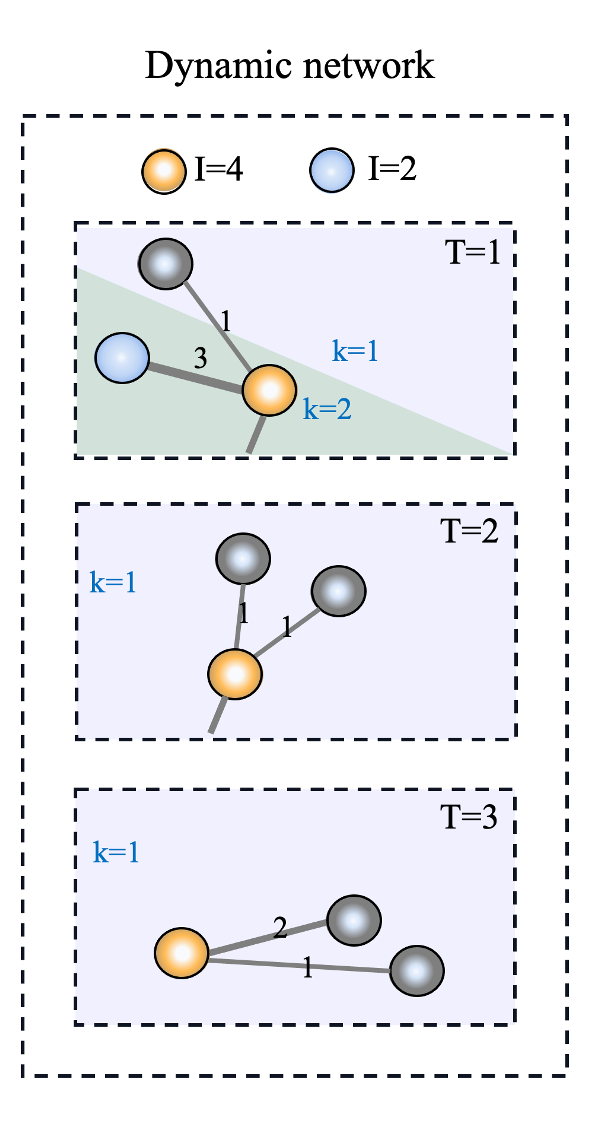}%
		\label{fig_kshell2}}
	\caption{Weighted k-shell decomposition. (a) on a static network, (b) extended to a dynamic network }
	\label{fig_kshell}
\end{figure}

Generally speaking, in a static network, the hubs are the key players when networks have a broad degree distribution. In addition, the topology of the network is also an important measure \cite{kitsak2010identification}. 
With the weighted k-shell decomposition method (W-KS) proposed in \cite{garas2012k}, we shall be able to rank the nodes according to both the degree and position of the node in each static undirected weighted network. Define the weighted degree $D_{t}^{i}$ of node $i$ in time frame $t$ as
\begin{equation}
	D^{i}_{t}=\left[\sqrt{d^{i}_{t}\cdot\sum_{j\in {n_t^i}}w^{ij}_{t}}\right],
	\label{eq_wd}
\end{equation}
the combination of degree $d^{i}_{t}$ and the sum of all its link weights $\sum_{j}w^{ij}_{t}$, rounded to the nearest integer. 
In W-KS, all nodes with $D$ not greater than 1 are removed first. Then, the $D$ of other nodes is recalculated in the trimmed network and the pruning process is repeated until no nodes with $D$ less than or equal to 1 are left in the network. The pruned nodes are grouped in the first shell with $k=1$.
Then the next k-shell with $k=2$ and further higher k-shells are separated from the remaining network iteratively until no nodes remain.
Finally each node has a value $k$, with larger $k$ indicating greater node influence. 
The influence of node $i$ in the $t$-th time frame, $I_{t}^{i}$, can be represented by 
\begin{equation}
	I_{t}^{i}=
	\left\{
	\begin{array}{l}
		k_{t}^{i},\; i\; in\; the\; t^{th}\; network \\
		0, \; i\; not\; in\; the\;t^{th}\; network
	\end{array}
.
	\right.
	\label{eq_coreness}
\end{equation}
The influence of node $i$ in the dynamic network is defined as the sum of its influence in each time frame
\begin{equation}
	I^{i}=\sum_{t=0}^{T}I_{t}^{i},
	\label{eq_influence}
\end{equation}
where $I_{t}^{i}$  is derived from the methods of static networks, i.e., in our case, weighted k-shell decomposition.

In Fig.~\ref{fig_kshell}, we show a particular example in which \subref{fig_kshell1} W-KS is applied to a static (but accumulated) network, and \subref{fig_kshell2} is extended and then applied to a dynamic network. It can be seen that the dynamic W-KS can better capture the temporal nature of the CSO and can thus characterize node influences more accurately.

\subsubsection{Community structure detection}
Backbone members (BMs) are identified as the top X\% of influential members, measured by any metric. We use the dynamic W-KS algorithm to calculate node influence, specifically coreness centrality, to select BMs. All other members are classified as general members (GMs). In turn, the network can be divided into three components: a) The sub-network formed by BMs and the interactions between them (BSN); b) The sub-network formed by GMs and the interactions between them (GSN); and c) The links that connect BSN and GSN.  
We use the quality metric modularity Q defined in \cite{blondel2008fast} to explore the community structure in the two sub-networks of the CSO respectively:
\begin{equation}
	Q_{t}=\dfrac{1}{2m}\sum_{i,j}{\left[w^{ij}_{t}-\dfrac{e_{t}^{i}e_{t}^{j}}{2m}\right]\delta\left(i, j\right)},
	\label{eq_modularity}
\end{equation}
where $e^{i}_{t}=\sum_j{w^{ij}_{t}}$ is the sum of the weights of the edges attached to node $i$,  and $m=\dfrac{1}{2}\sum_{ij}{w_{t}^{ij}}$ is the sum of the weights of all edges in $G_{t}$. The $\delta$-function is 1 if node $i$ and $j$ in the same community, otherwise $\delta=0$. A $Q$ value higher than 0.3 suggests that distinct community structures do exist in the network \cite{2009Mining}. 

\begin{figure*}[t]
	\centerline{\includegraphics[width=5in]{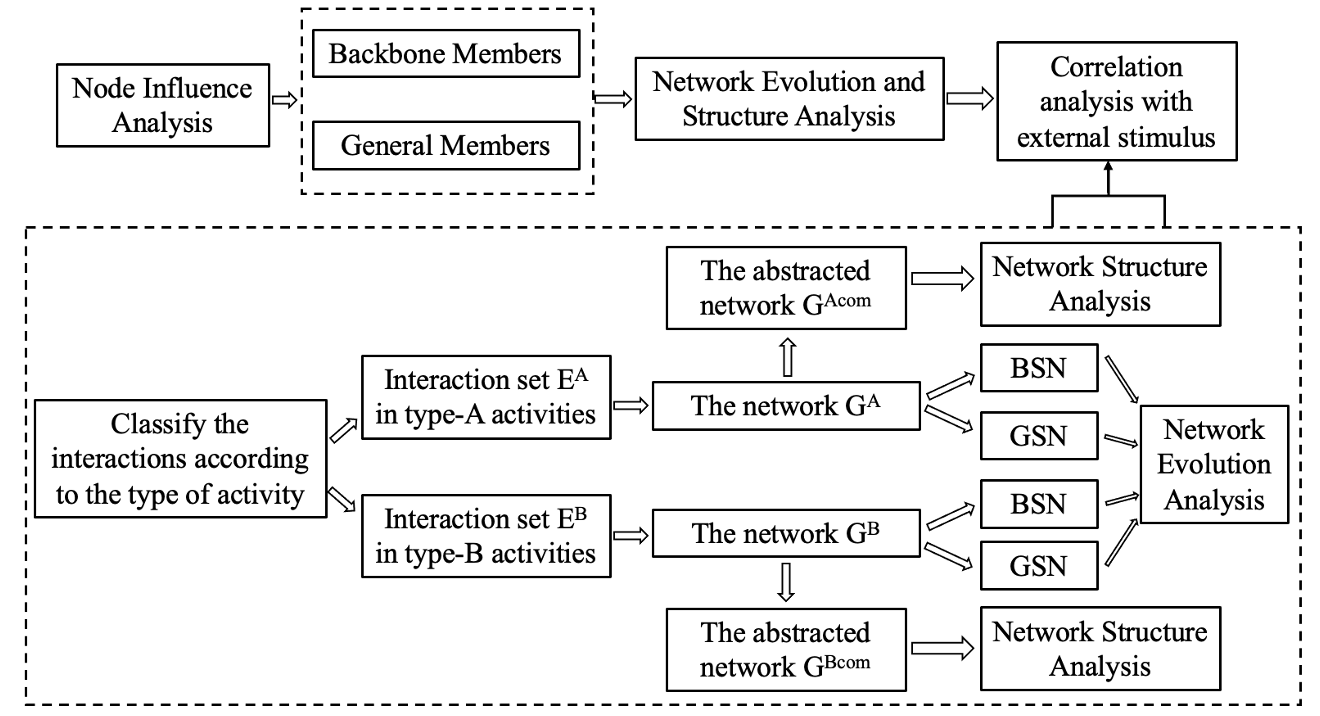}}
	\vspace{-3mm}
	\caption{Explanation of specific experimental steps.}
	\label{fig_exp}
\end{figure*}

\subsection{Tier-two Analysis}

\subsubsection{Community evolution}
Once the presence of community structure is confirmed, we can then proceed with the analysis on community evolution. To capture intermittent participation that is often seen in CSOs, we extend the community evolution events used in our previous study  \cite{yu2021community}, by adding $Suspend$ and $\mathit{Re \mhyphen emerge}$. A community is said to be suspended, if it appears in a time frame, but disappears for some time, and then re-emerges. A community is said to be re-emerging, if it did not exist in the previous time frame, but has appeared at least once in the past time frames. According to their effects on the community structure, the evolution events other than $Continue$ may be roughly classified into two categories: a) Events that bring significant structural changes to the network ($V$ - Violent), mainly through adding/removing nodes from the network - $Form$, $Dissolve$, $Suspend$ and $\mathit{Re \mhyphen emerge}$ and b) Events that cause marginal structural changes to the network ($S$ - Stable), mainly through adding/removing links between existing nodes - $Grow$, $Merge$, $Shrink$ and $Split$.  Communities that undergo $S$-type evolution are relatively closed groups with close internal interactions but limited communication with other external communities. However, $V$-type evolution brings about more diverse changes in the network structure, which promotes communication among different groups and plays an important role in the stability and development of the network. The information about community evolution events is presented in Table \ref{tab_event} in detail.

\subsubsection{Community abstraction}
To explore the interactions of different communities on a horizontal level, we abstract the network of communities by hiding the details of the original connections between individuals within communities. The network described by Eq.(\ref{eq_GraphCom}) is similar to Eq. (\ref{eq_Graph}), however, the nodes are now communities detected in Tier-one and edges are connections between communities.  
 \begin{equation}
	G^{com}=\{G_{t}^{com}=(V^{com}_{t}, E^{com}_{t}), \ for\ all\ t\ in\ [0, T]\},
	\label{eq_GraphCom}
\end{equation}

There are two kinds of nodes in the network: a) communities formed by backbone members (BCs); and b) communities formed by general members (GCs). The connections in the network are classified into three categories: a) edges between BCs (BBEs); b) edges between GCs (GGEs); and c) edges between BCs and GCs (BGEs).

We characterize the structure of the network using network density and betweeness centrality, where network density is denoted as 
 \begin{equation}
	Density(G^{com})=\frac{2L^{com}}{N^{com}(N^{com}-1)},
	\label{eq_density}
\end{equation}
where $N^{com}$ denotes the number of communities in the network and $L^{com}$ denotes the number of connected edges between communities in the network.
The betweeness centrality is formulated as
\begin{equation}
	BC^{com}_{z}=\sum_{m,n \in V^{com}}{\frac{\sigma(m, n|z)}{\sigma(m, n)}},
	\label{eq_between}
\end{equation}
where $m$ and $n$ denote any community.

The network with core-periphery structure has a high backbone node centrality and a high density of BSN. However, it is difficult to form a connected network by connections between general nodes only.

\section{Experiments and Results}
In this section, we apply the proposed Twotier framework to a sports organization and obtain information about the evolution of the organization and its structural characteristics. The experimental steps are illustrated in Fig. \ref{fig_exp}.

The data that we use are collected from a non-profit sports organization, through an online platform serving a community with more than 10,000 members. Users can organize communal activities, most of which require users to participate in teams with less than 10 members. Necessary tools are provided for team members to communicate with each other. The platform went online in May 2015, and by June 2021, 790 activities have been held, with 4879 different individual participants in 6426 teams. The activities can be rewarding, denoted as Type-A, with a total of 119 activities, or without any reward, with a total of 671, denoted as Type-B.

According to the teaming relations in the entire time span,  each different user is represented with a node, and team-wise relations are represented by undirected links with weight of 1. It is clear that a link is in fact representing the quadruple of $\left\langle node \ pair,\ id\ of\ team,\ id\ of\ the\ activity,\ time\ of\ the\ \mathit{act \mhyphen }\notag\right. \\ \left.ivity,\ type\ of\ the\ activity\right\rangle$. In the entire time span, there are 73813 such links.

\begin{figure}[t]
	\centering
	\includegraphics[width=0.45\textwidth]{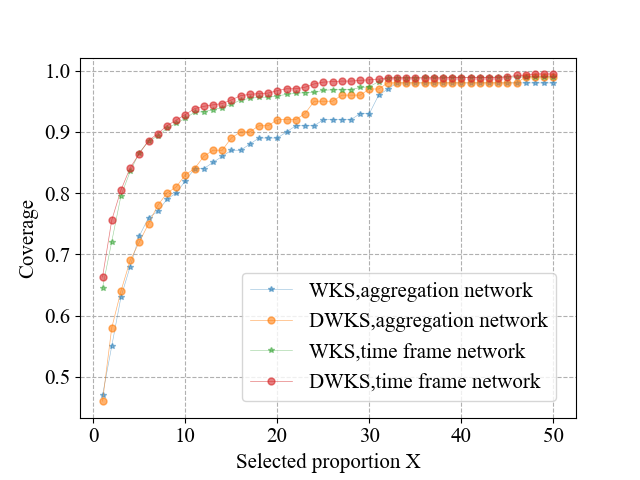}
	\caption{Coverage comparison between weighted k-shell decomposition (WKS) and dynamic weighted k-shell decomposition (DWKS) methods in two network formats.}
	\label{fig_coverage}
\end{figure}
\subsection{Influence of Nodes by dynamic weighted k-shell decomposition (dynamic W-KS)}

While traditional weighted k-shell decomposition (W-KS) divides all nodes into 87 shells, dynamic W-KS divides all nodes into 457 shells, allowing for a more detailed division with more prominent gaps between nodes. Furthermore, for nodes within the same layer, we sort them based on their degree. To verify that the extended node influence determination approach is superior, we compare the coverage when the top X\% members are selected as backbone members. The coverage reflects the range of influence in the network. It is defined as the proportion of the number of selected kernel members, and their neighbors, to the size of the given network. 

We compare in two network scenarios, one ignoring temporal properties and aggregating all members who have ever appeared in the organization to form an aggregation network; the other creates a time-frame network at three-month intervals. As shown in Fig. \ref{fig_coverage}, with selected proportion $X$ increases from 1 to 50, the coverage under both methods increases but dynamic W-KS generally gives higher results than W-KS.  Considering degree, the position in network topology and activeness, the dynamic W-KS can give a more comprehensive picture of the importance of a node and is therefore chosen in our further analysis.
In the following parts, we choose the cases $X = 5$, $X=10$, and $X=20$ to carry out the experiment respectively. 

After classifying the network members, we find that the BMs have a greater degree and are in a more important position in the network topology than GMs, as the results in Table \ref{tab_backbone} show.  In addition, they are involved in a larger number of activities and active in the network for a longer period of time than GMs. On average, the participation of BMs in non-rewarding activities is higher than in rewarding activities, which is the opposite of GMs. The network structure formed by each of these two groups is also different.

\subsection{The Evolution of communities}

\begin{table}[]
	\caption{Properties of BMs and GMs}
	\begin{center}
		\begin{tabular}{c|cc|cc|cc}
			\toprule[1pt]
			\textbf{Proportion}                                                                    & \multicolumn{2}{c|}{\textbf{5\%}}                & \multicolumn{2}{c|}{\textbf{10\%}}               & \multicolumn{2}{c}{\textbf{20\%}}                \\ \hline
			\textbf{Properties}                                                                           & \multicolumn{1}{c|}{\textbf{BMs}} & \textbf{GMs} & \multicolumn{1}{c|}{\textbf{BMs}} & \textbf{GMs} & \multicolumn{1}{c|}{\textbf{BMs}} & \textbf{GMs} \\ \hline
			\textbf{Avg. degree}                                                                   & \multicolumn{1}{c|}{130}          & 17           & \multicolumn{1}{c|}{92}           & 14           & \multicolumn{1}{c|}{60}           & 10           \\
			\textbf{\begin{tabular}[c]{@{}c@{}}Avg. closeness \\ centrality\end{tabular}}          & \multicolumn{1}{c|}{0.4}          & 0.3          & \multicolumn{1}{c|}{0.4}          & 0.3          & \multicolumn{1}{c|}{0.4}          & 0.3          \\
			\textbf{\begin{tabular}[c]{@{}c@{}}Avg.  participation \\ in type-A\end{tabular}} & \multicolumn{1}{c|}{29}           & 2.3          & \multicolumn{1}{c|}{20}           & 1.8          & \multicolumn{1}{c|}{12}           & 1.4          \\
			\textbf{\begin{tabular}[c]{@{}c@{}}Avg. participation \\ in type-B\end{tabular}} & \multicolumn{1}{c|}{54}           & 0.5          & \multicolumn{1}{c|}{30}           & 0.3          & \multicolumn{1}{c|}{15}           & 0.2          \\
			\textbf{Avg. active frames}                                                      & \multicolumn{1}{c|}{14}           & 2.3          & \multicolumn{1}{c|}{10}           & 2.0          & \multicolumn{1}{c|}{7.6}          & 1.7         \\ \bottomrule[1pt]
		\end{tabular}
		\label{tab_backbone}
	\end{center}
\end{table}

\begin{table}[]
	\caption{Sub-network modularity in different cases of backbone ratio}
	\begin{center}
	\begin{tabular}{c|cc|cc|cc}
		\hline
		\textbf{Proportion}  & \multicolumn{2}{c|}{\textbf{5\%}}                & \multicolumn{2}{c|}{\textbf{10\%}}               & \multicolumn{2}{c}{\textbf{20\%}}                \\ \hline
		\textbf{Sub-network} & \multicolumn{1}{c|}{\textbf{BSN}} & \textbf{GSN} & \multicolumn{1}{c|}{\textbf{BSN}} & \textbf{GSN} & \multicolumn{1}{c|}{\textbf{BSN}} & \textbf{GSN} \\ \hline
		\textbf{Modurality}  & \multicolumn{1}{c|}{0.40}         & 0.75         & \multicolumn{1}{c|}{0.43}         & 0.86         & \multicolumn{1}{c|}{0.47}         & 0.85         \\ \hline
	\end{tabular}
	\end{center}
\end{table}

\begin{figure*}[t]
	\centering  
	\subfigbottomskip=2pt 
	\subfigure[Percentage of nine evolutionary events while X=5]{
		\includegraphics[width=0.93\textwidth]{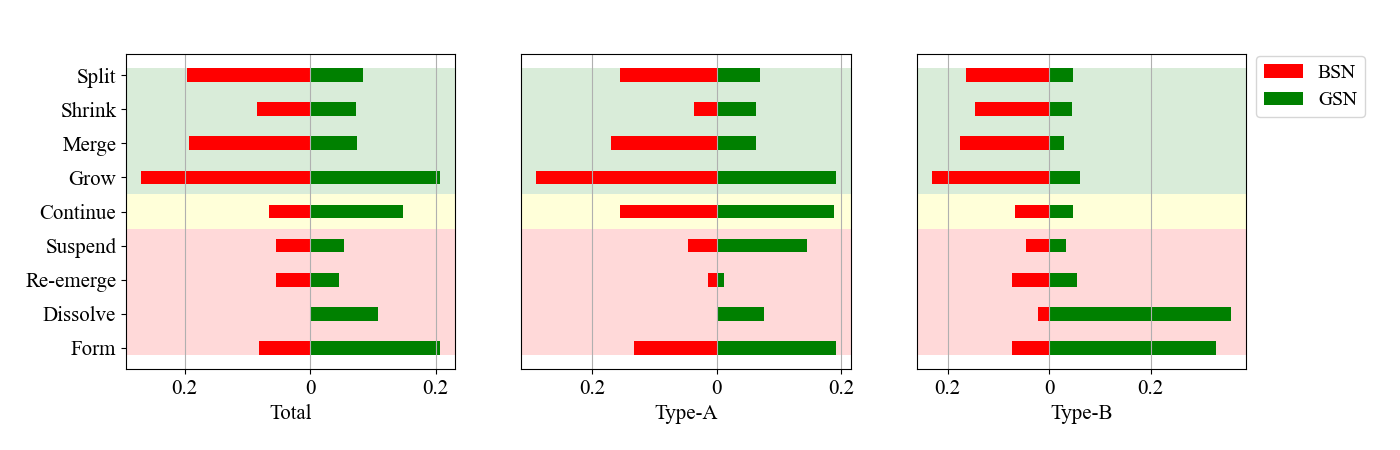}}
	\subfigure[Percentage of nine evolutionary events while X=10]{
		\includegraphics[width=0.93\textwidth]{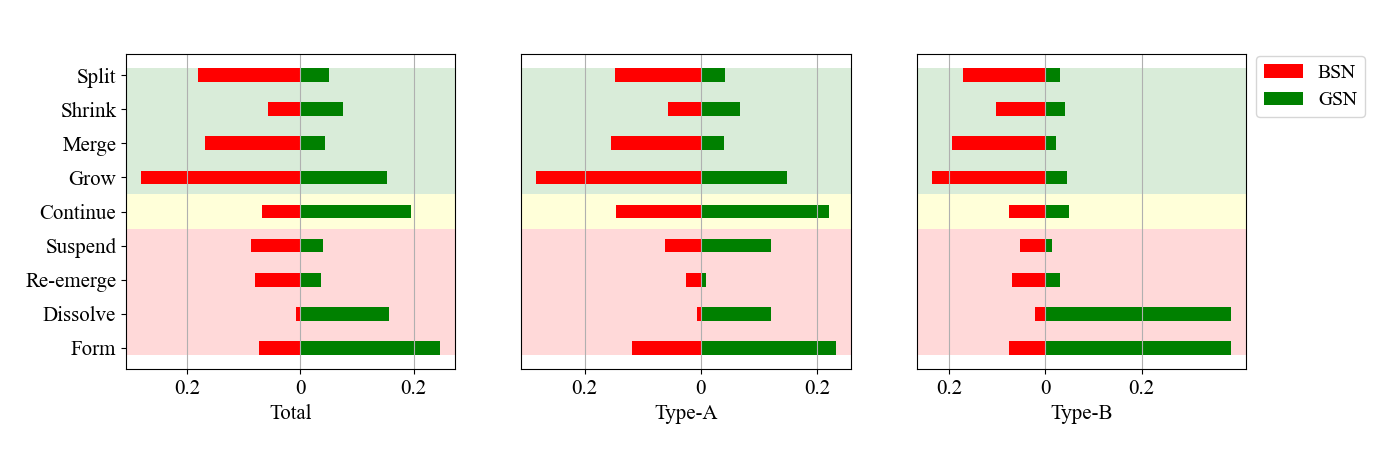}}
	\subfigure[Percentage of nine evolutionary events while X=20]{
		\includegraphics[width=0.93\textwidth]{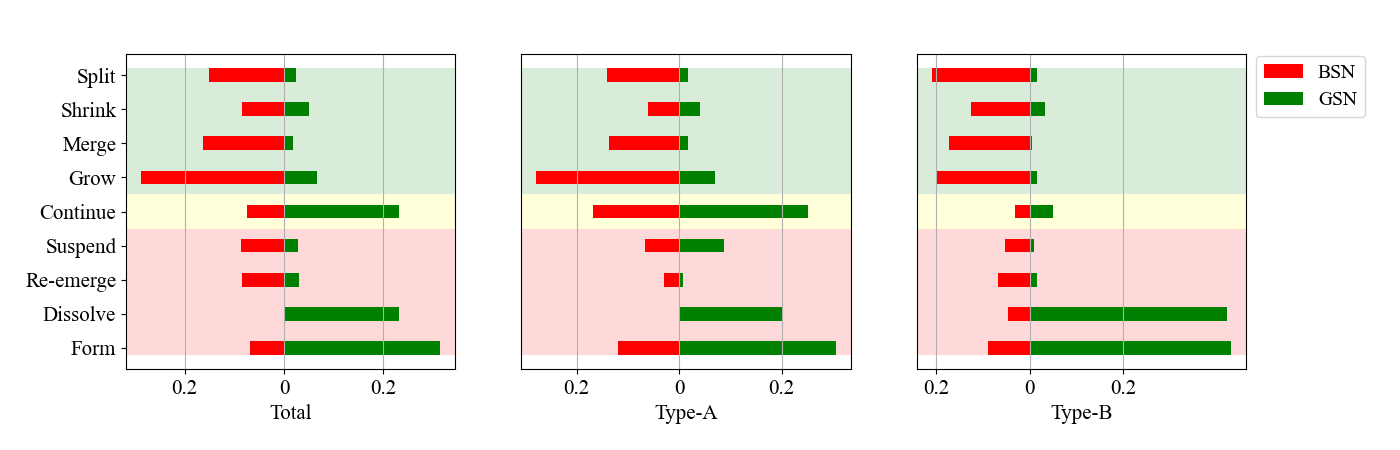}}
	\caption{Comparison of community evolution events in sub-networks when the percentage of backbone members varies.}
	\label{figEventCount}
	\vspace{-2mm}
\end{figure*}

In each time frame, the BSN has a giant component, occasionally there are also some small groups independent of the huge part, while the GSN is consist of many small groups that are separate from each other. The average $Q$ of the two sub-networks across all time frames is higher than 0.3, suggesting that they both have a distinct community structure. 

Further, we present the evolutionary relationships of the detected communities after dividing the 9 evolution events into three categories. Fig.~\ref{figEventCount} illustrates the percentage of evolution events for BSN and GSN, with rewarding activities (type-A), with non-rewarding activities (type-B), and with both, respectively. It can be observed that in GSN under type-B activities, $form$ and $dissolve$ among these 9 evolution events are always the major ones. However, this is very different in GSN under type-A activities with diverse and abundant events. With no stimulus, the GMs are less willing to participate, resulting in a large number of dropouts after attending the activities. This suggests that type-B activities are suitable to be held regularly to consolidate the connection between BMs rather than to absorb new members. It can also be seen that the BSN has more $S$-type community evolution events  (light green area), while there is a higher number of $V$-type community evolution events  (light red area) in GSN. The mobility within the backbone member groups is stronger than that of general members. This indicates that the backbone groups play the role of the trunk, while other members renew at a faster rate and act like leaves.

\subsection{The Structure of Abstracted Network}
\begin{figure*}[t]
	\centering  
	\subfigbottomskip=2pt 
	\subfigure[slot0]{
		\fbox{\includegraphics[width=0.146\linewidth, height=0.14\linewidth]{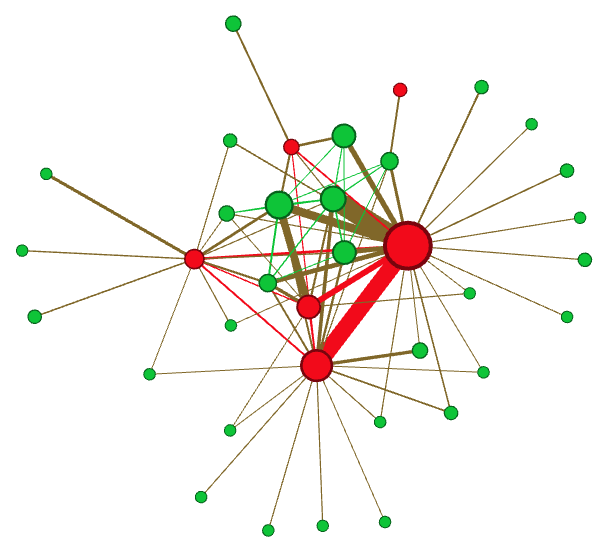}}}
	\hspace{-4mm}
	\subfigure[slot4]{
		\fbox{\includegraphics[width=0.146\linewidth, height=0.14\linewidth]{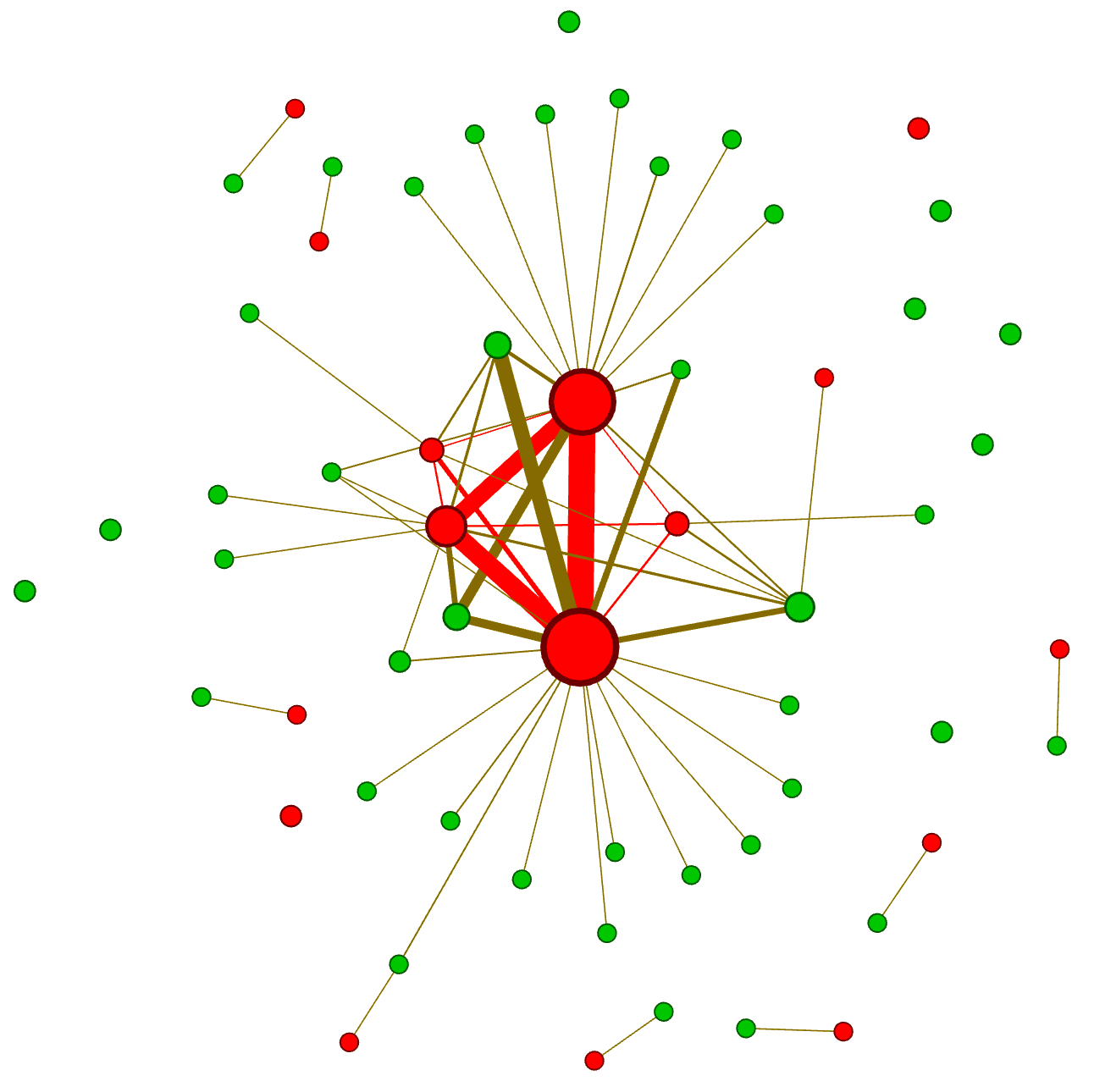}}}
	\hspace{-4mm}
	\subfigure[slot8]{
		\fbox{\includegraphics[width=0.146\linewidth, height=0.14\linewidth]{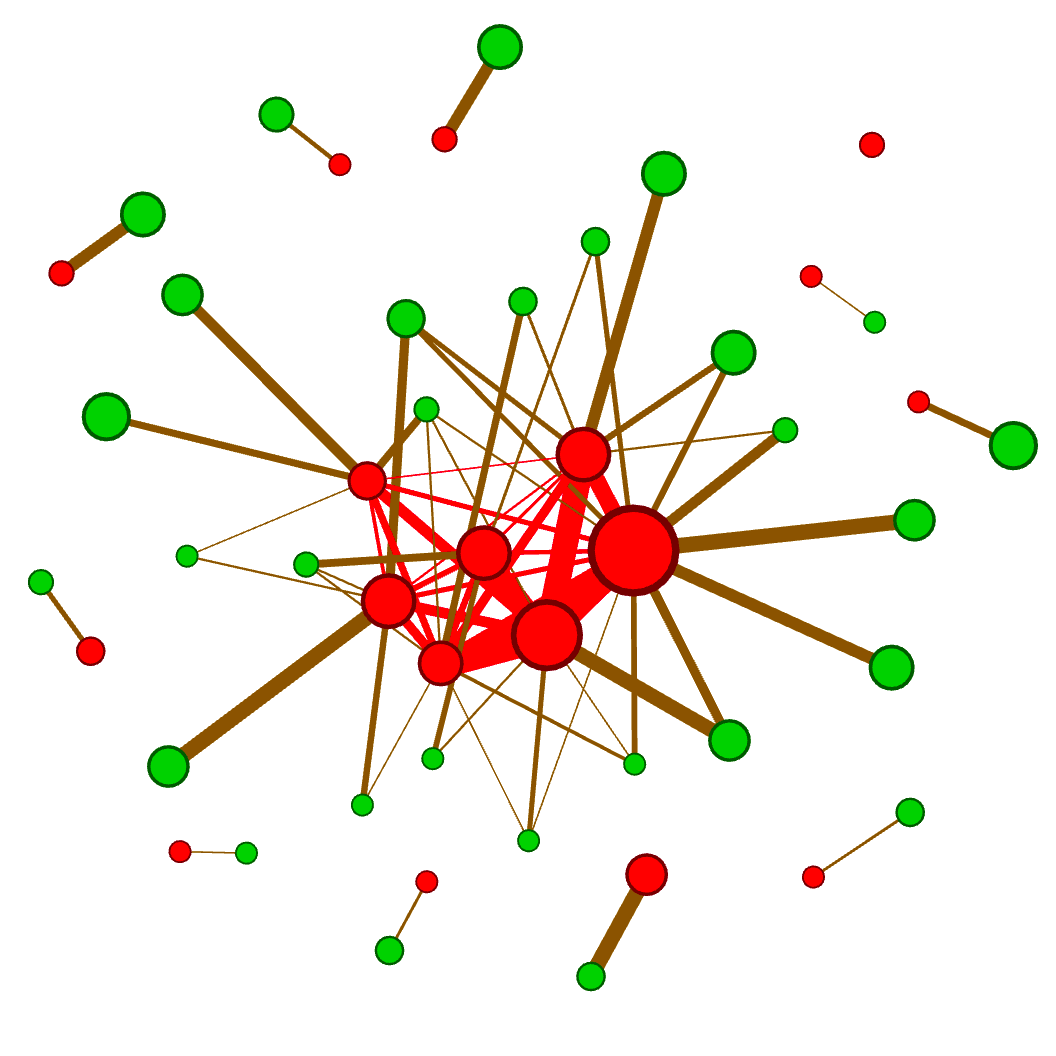}}}
	\hspace{-4mm}
	\subfigure[slot12]{
		\fbox{\includegraphics[width=0.146\linewidth, height=0.14\linewidth]{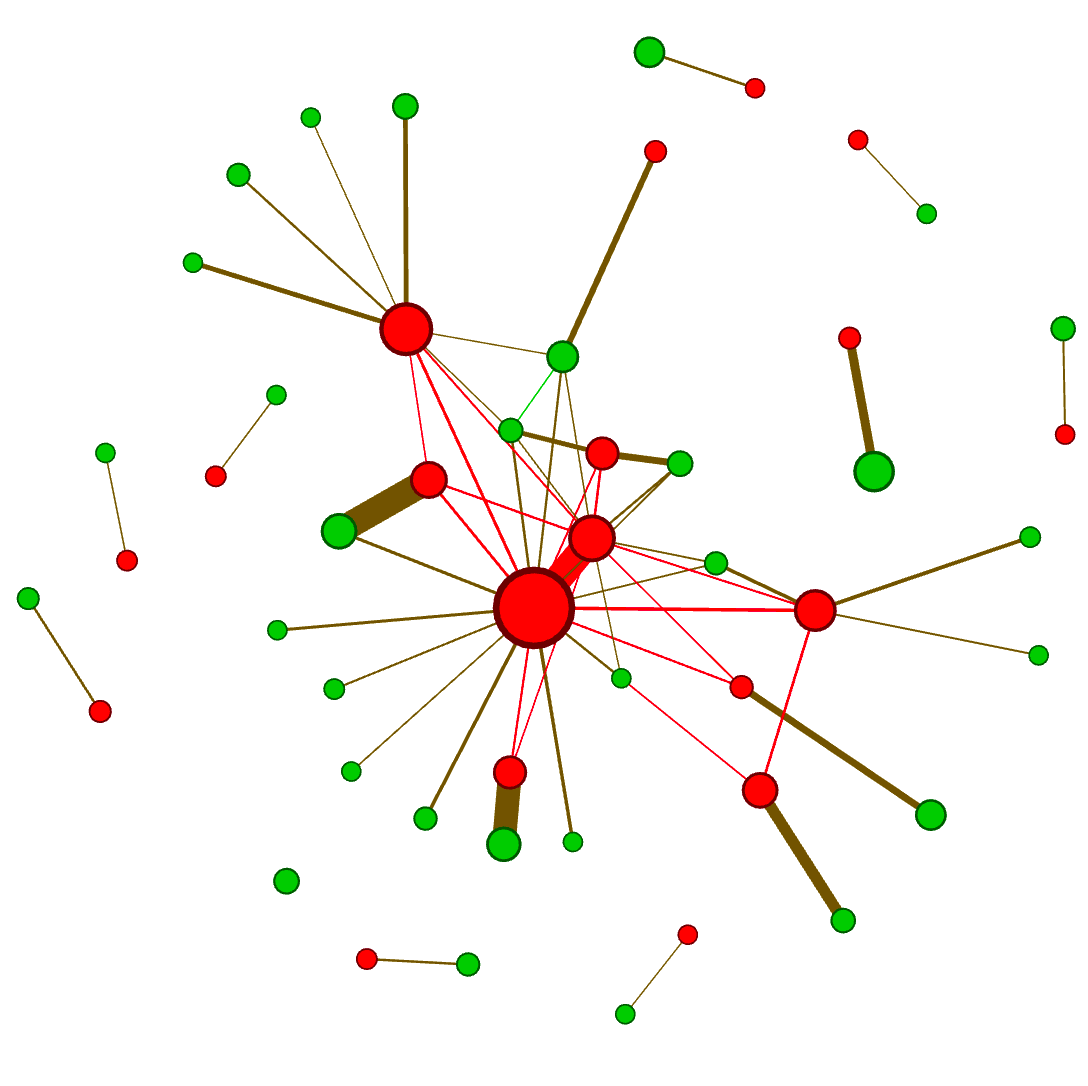}}}
	\hspace{-4mm}
	\subfigure[slot16]{
		\fbox{\includegraphics[width=0.146\linewidth, height=0.14\linewidth]{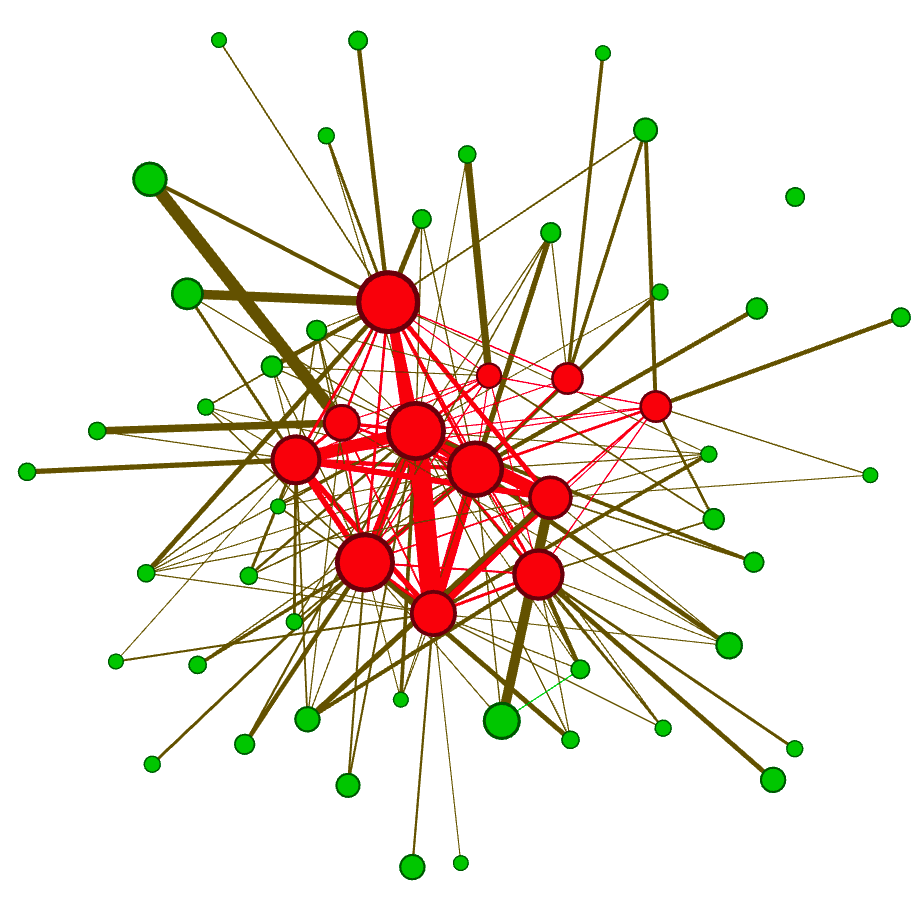}}}
	\hspace{-4mm}
	\subfigure[slot20]{
		\fbox{\includegraphics[width=0.146\linewidth, height=0.14\linewidth]{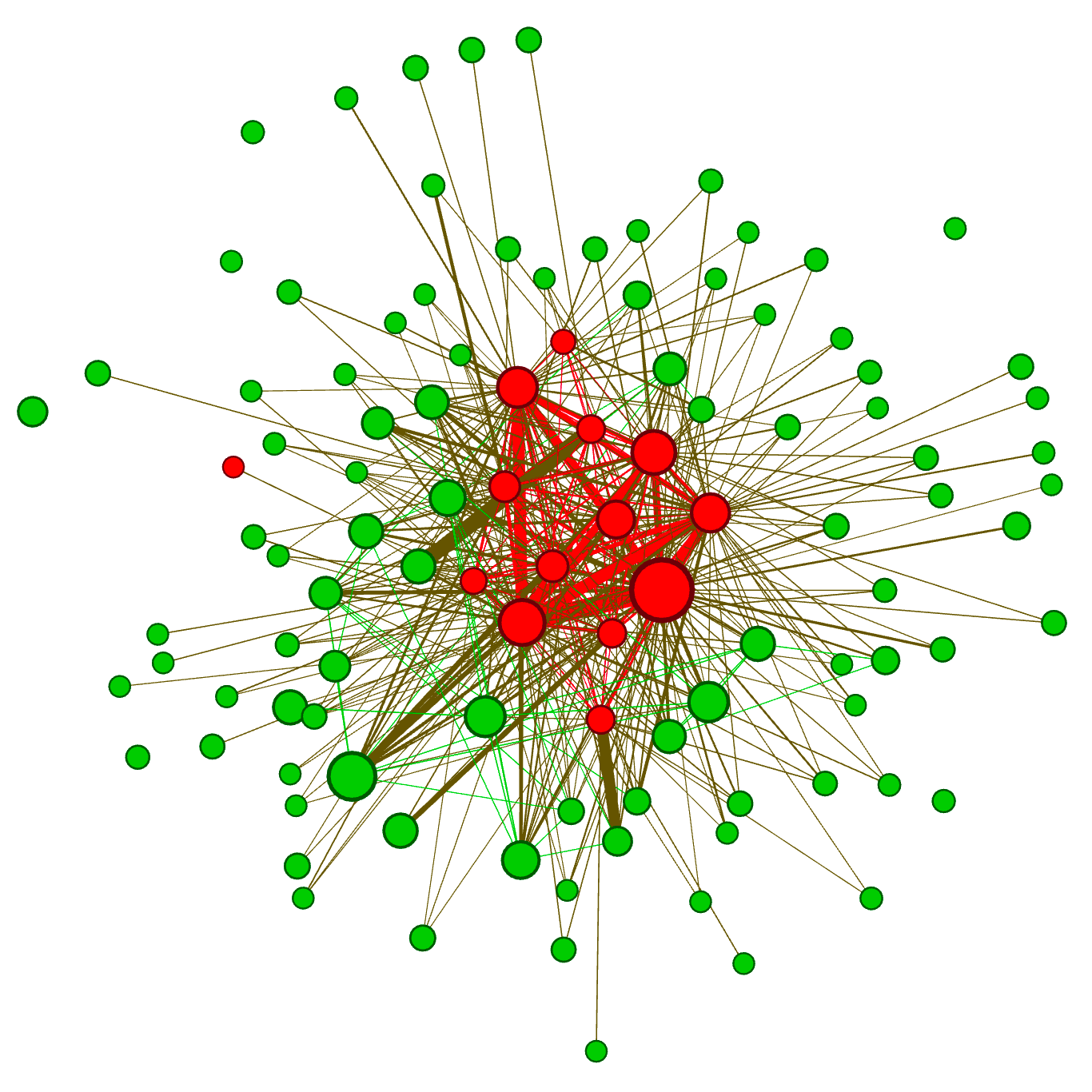}}}
	\caption{Network with community as node in some time frames while $X=10$.}
	\label{figCom}
	\vspace{-2mm}
\end{figure*}

\begin{figure*}[t]
	\centering
	\subfigure[]{\includegraphics[width=0.325\textwidth]{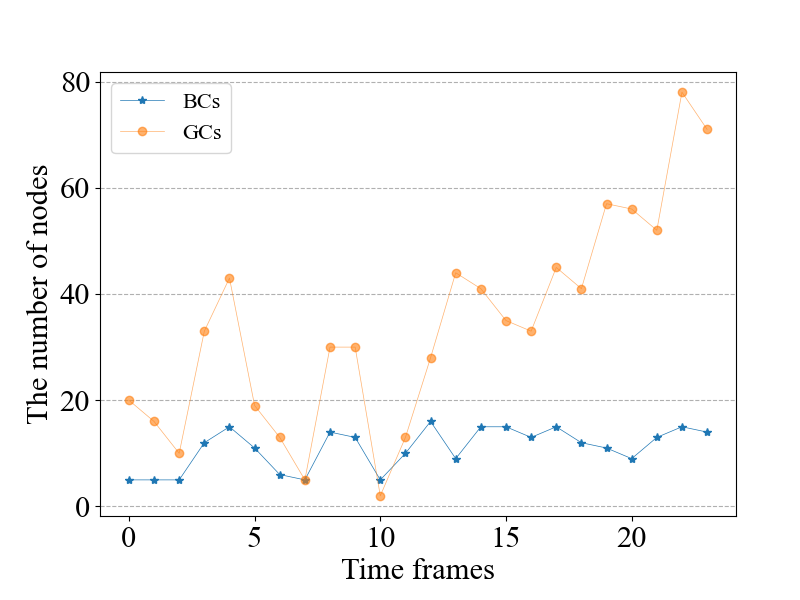}%
		\label{figcomnum5}}
	\hfil
	\subfigure[]{\includegraphics[width=0.325\textwidth]{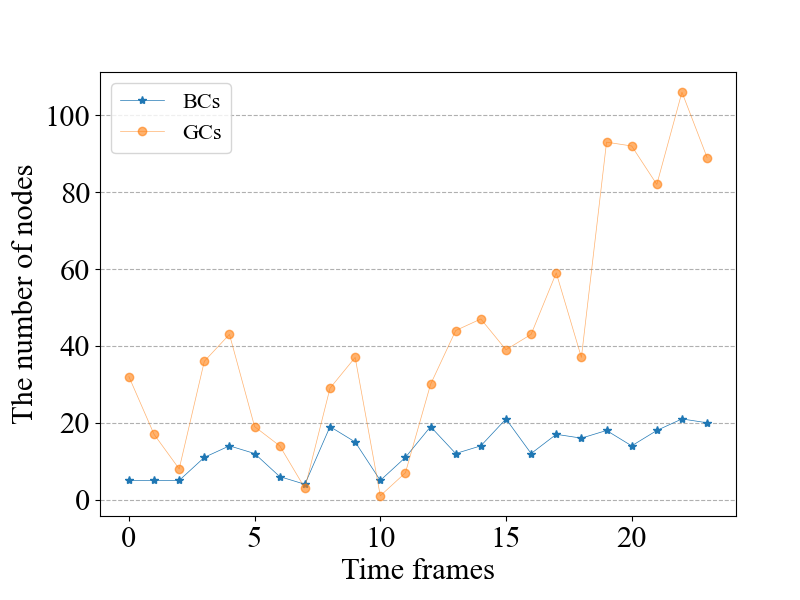}%
		\label{figcomnum10}}
	\hfil
	\subfigure[]{\includegraphics[width=0.325\textwidth]{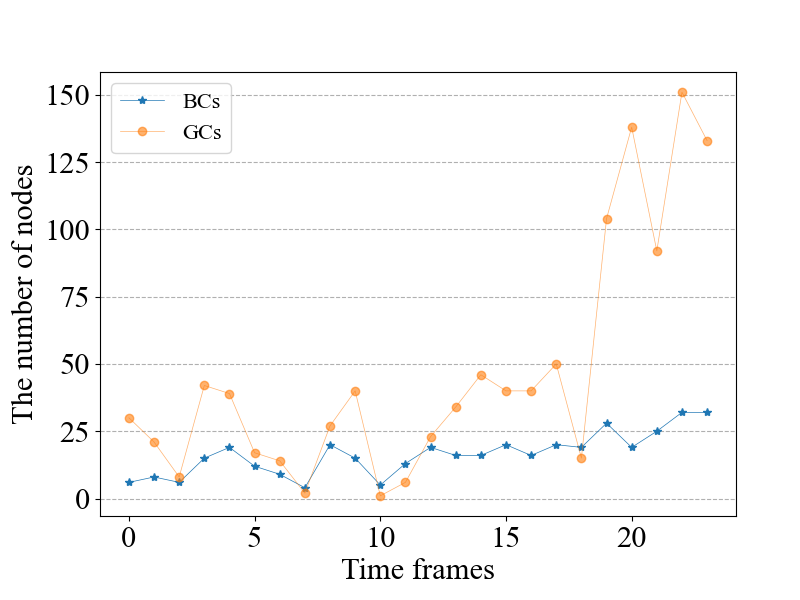}%
		\label{figcomnum20}}
	\hfil
	\subfigure[]{\includegraphics[width=0.325\textwidth]{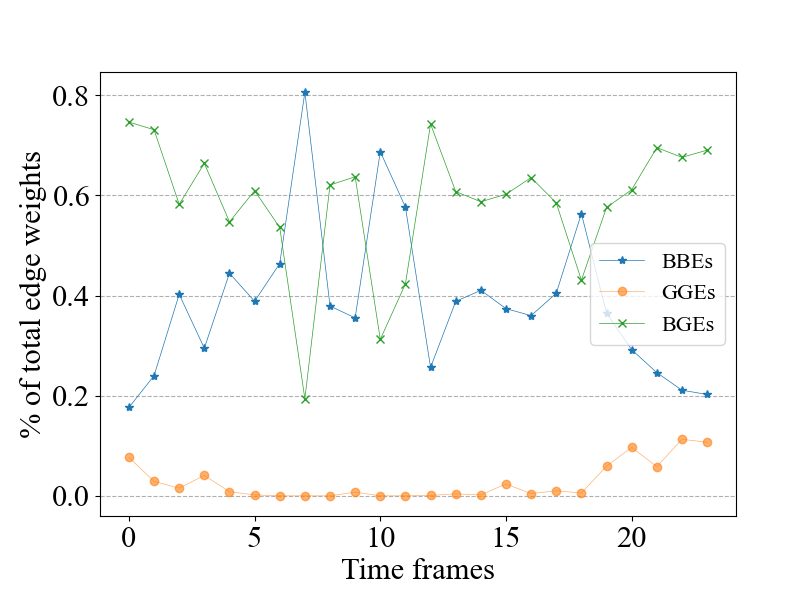}%
		\label{figedge5}}
	\hfil
	\subfigure[]{\includegraphics[width=0.325\textwidth]{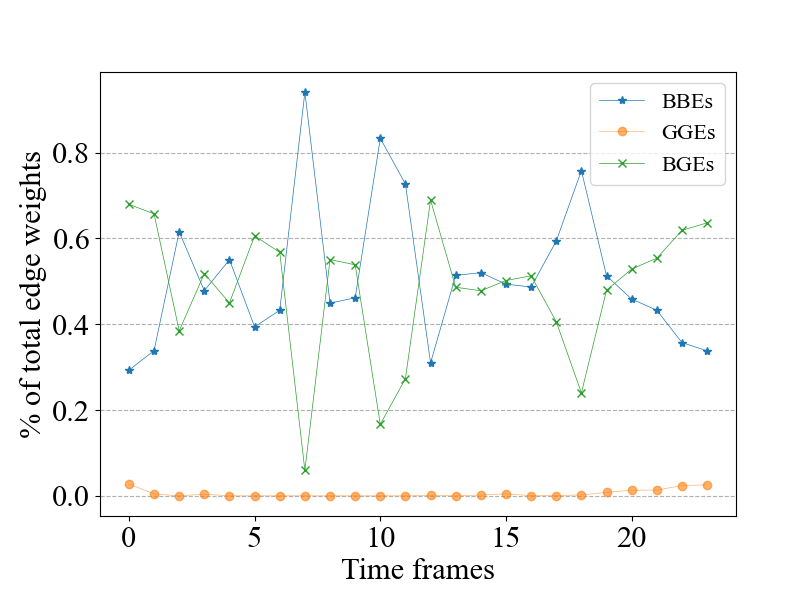}%
		\label{figedge10}}
	\hfil
	\subfigure[]{\includegraphics[width=0.325\textwidth]{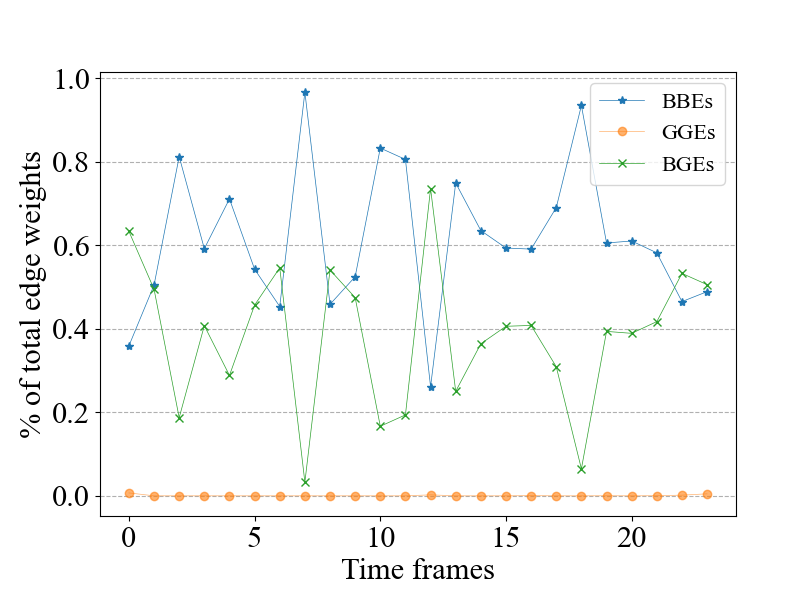}%
		\label{figedge20}}
	\caption{Network of communities. (a) The number of network nodes while X=5, (b) The number of network nodes while X=10, (c) The number of network nodes while X=20, (d) Percentage of the three kinds of edge weights while X=5, (e) Percentage of the three kinds of edge weights while X=10, (b) Percentage of the three kinds of edge weights while X=20.}
	\label{figComEdge}
\end{figure*}

\begin{figure*}[t]
	\centering
	\subfigure[]{\includegraphics[width=0.325\textwidth]{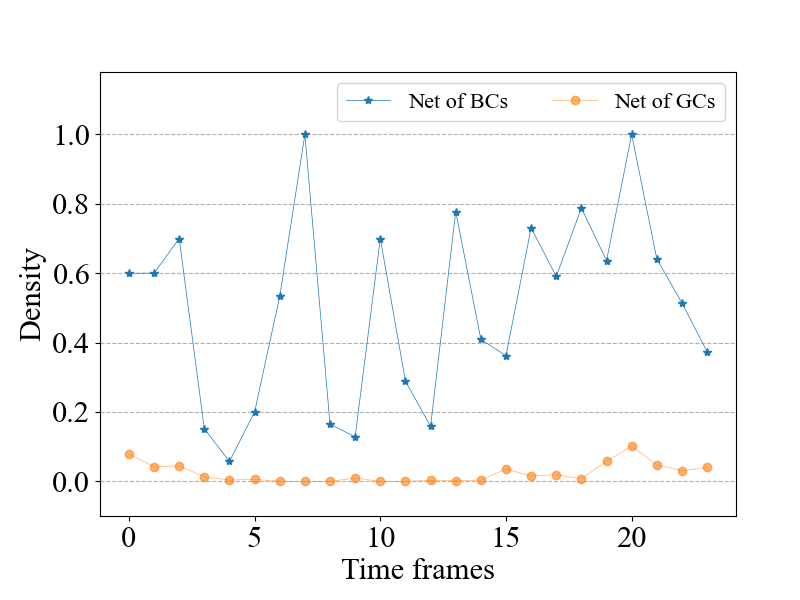}%
		\label{figdensity5}}
	\hfil
	\subfigure[]{\includegraphics[width=0.325\textwidth]{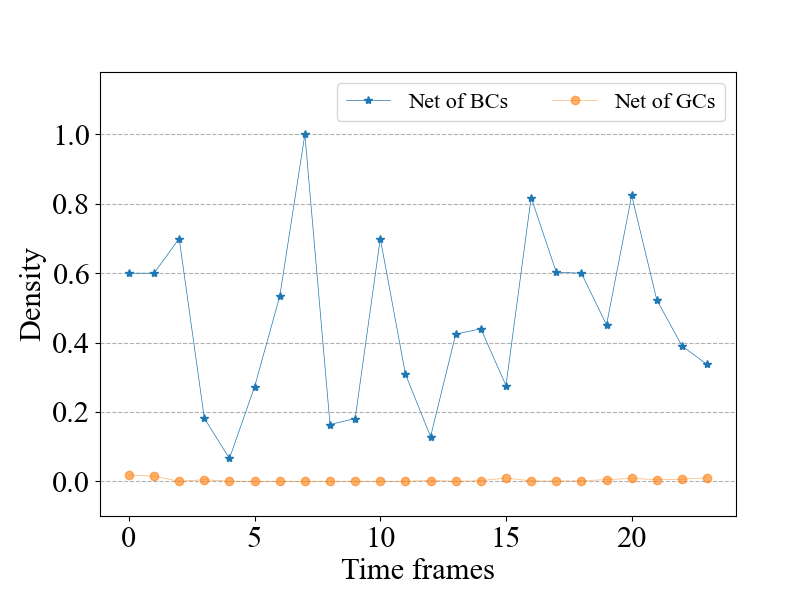}%
		\label{figdensity10}}
	\hfil
	\subfigure[]{\includegraphics[width=0.325\textwidth]{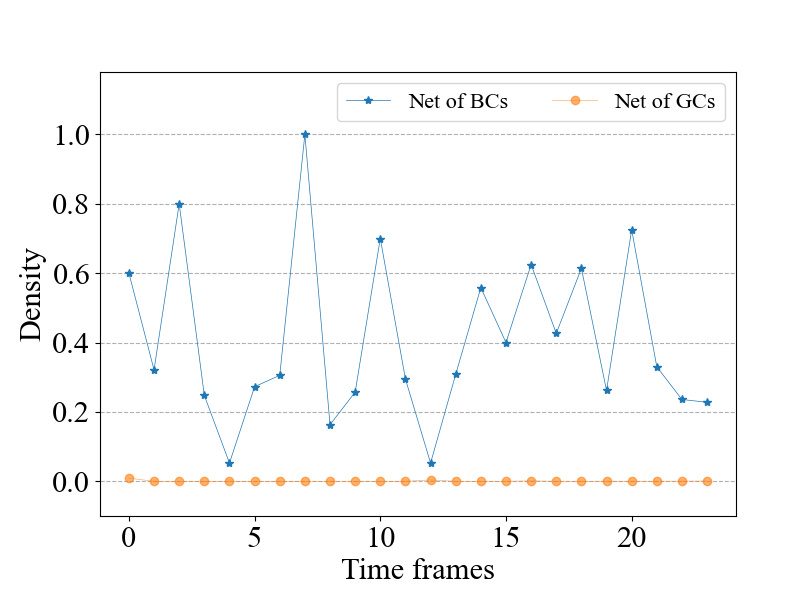}%
		\label{figdensity20}}

	\caption{Density of sub-networks. (a) X=5, (b) X=10, (c) X=20. }
	\label{figDensity}
\end{figure*}

\begin{figure*}[t]
	\centering
	\subfigure[]{\includegraphics[width=0.325\textwidth]{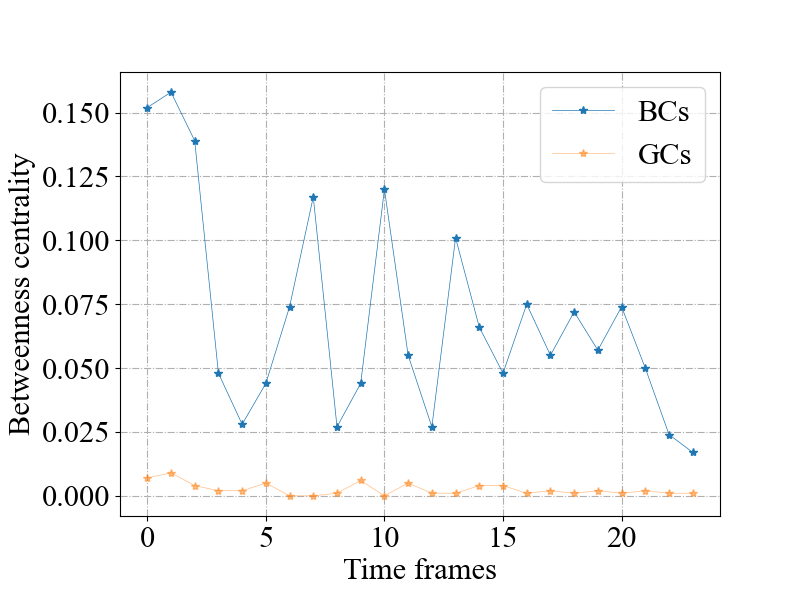}%
		\label{figbet5}}
	\hfil
	\subfigure[]{\includegraphics[width=0.325\textwidth]{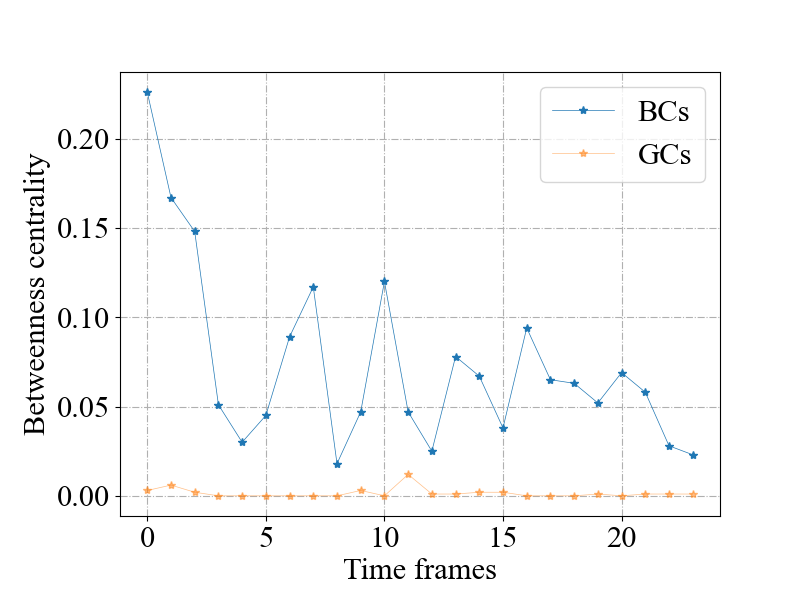}%
		\label{figbet10}}
	\hfil
	\subfigure[]{\includegraphics[width=0.325\textwidth]{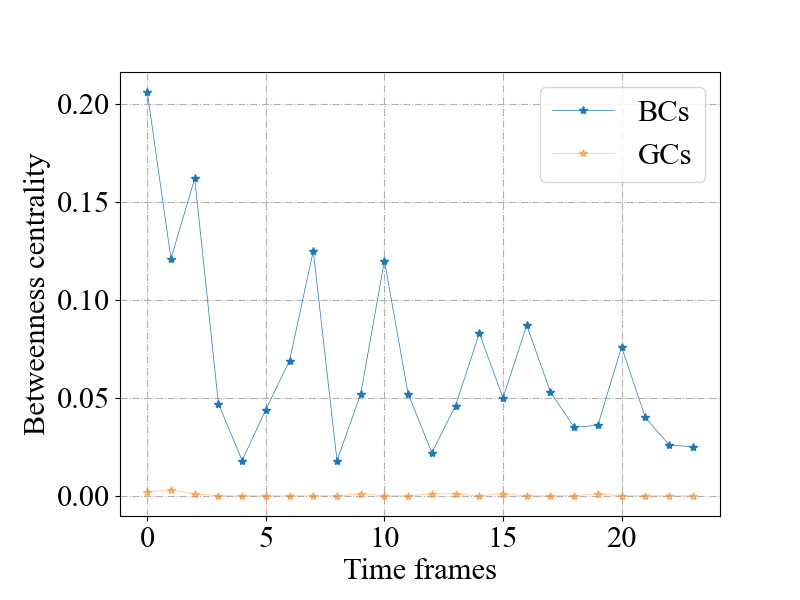}%
		\label{figbet20}}
	
	\caption{Betweeness centrality of BCs and GCs in abstracted network. (a) X=5, (b) X=10, (c) X=20. }
	\label{figBetweeness}
\end{figure*}

\begin{figure*}[t]
	\centering
	\subfigure[]{\includegraphics[width=0.45\textwidth]{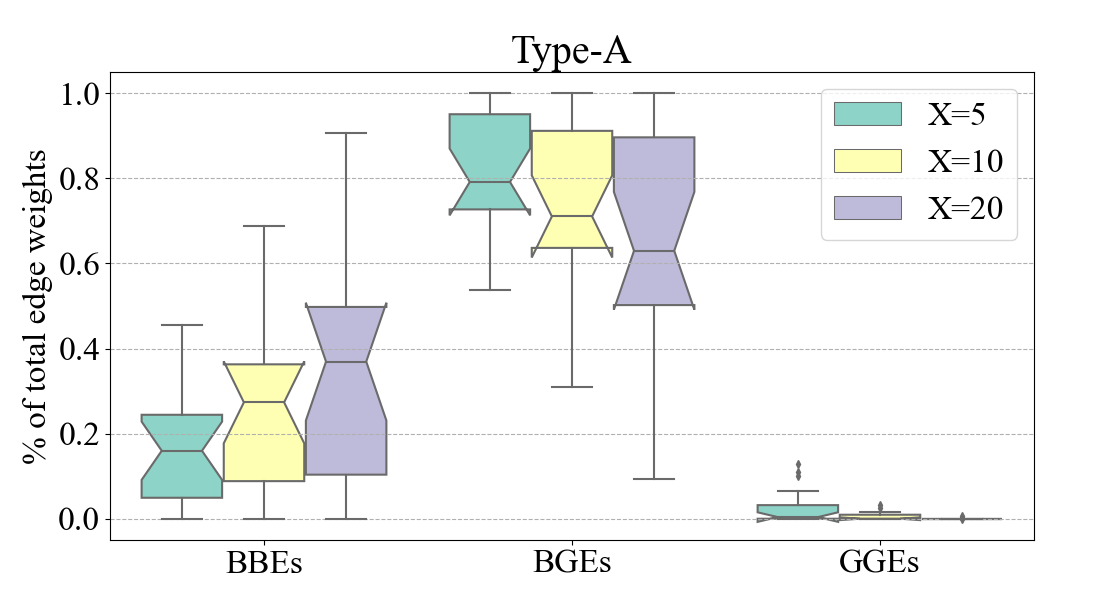}%
		\label{fig2edgeA}}
	\hfil
	\subfigure[]{\includegraphics[width=0.45\textwidth]{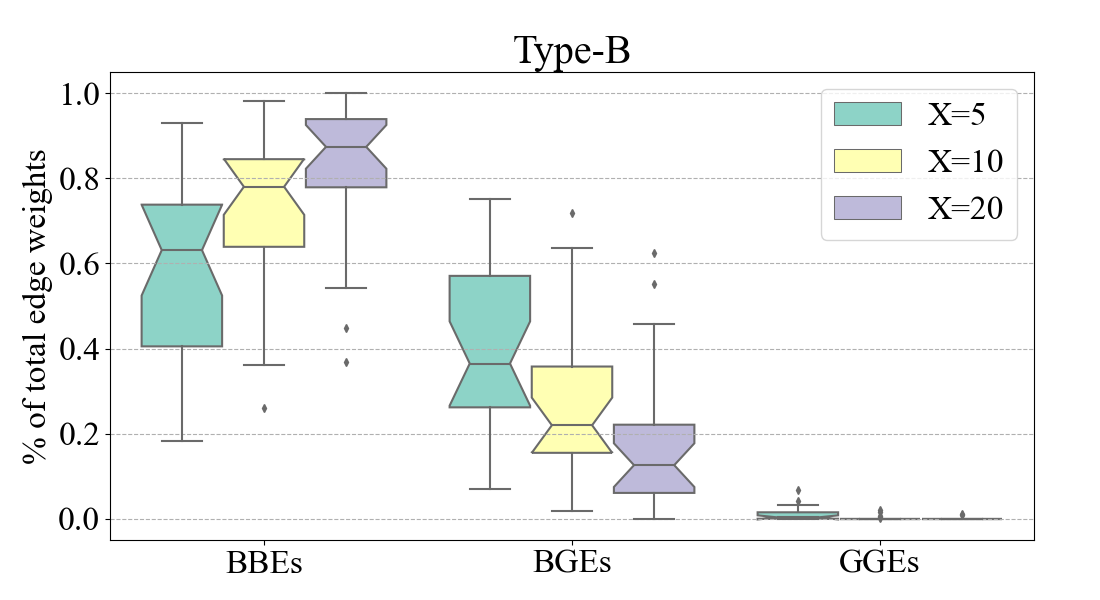}%
		\label{fig2edgeB}}
	\caption{Percentage of the three kinds of edge weights while $X$=5,10 and 20, (a) in the network $G^{A}$, (b)  in the network $G^{B}$. }
	\label{fig2edge}
\end{figure*}

\begin{figure*}[t]
	\centering
	\subfigure[]{\includegraphics[width=0.325\textwidth]{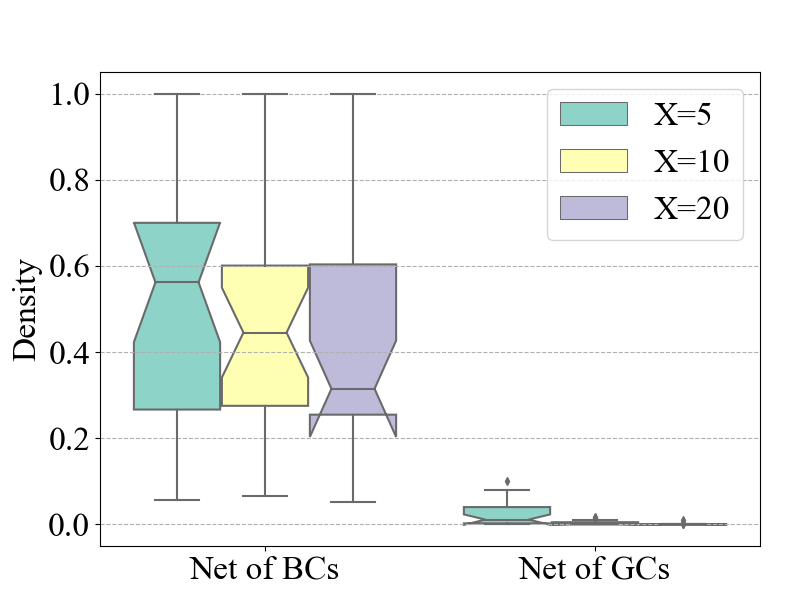}%
		\label{fig3density}}
	\hfil
	\subfigure[]{\includegraphics[width=0.325\textwidth]{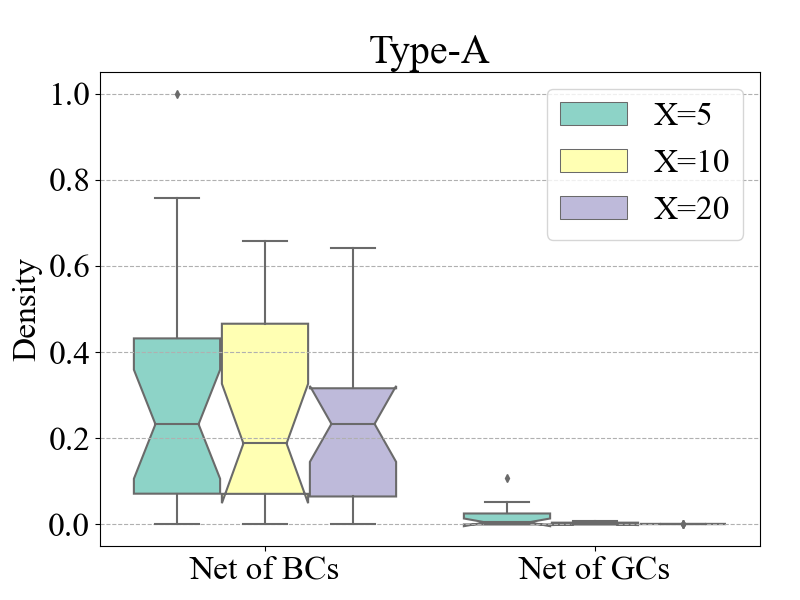}%
		\label{fig3densityA}}
	\hfil
	\subfigure[]{\includegraphics[width=0.325\textwidth]{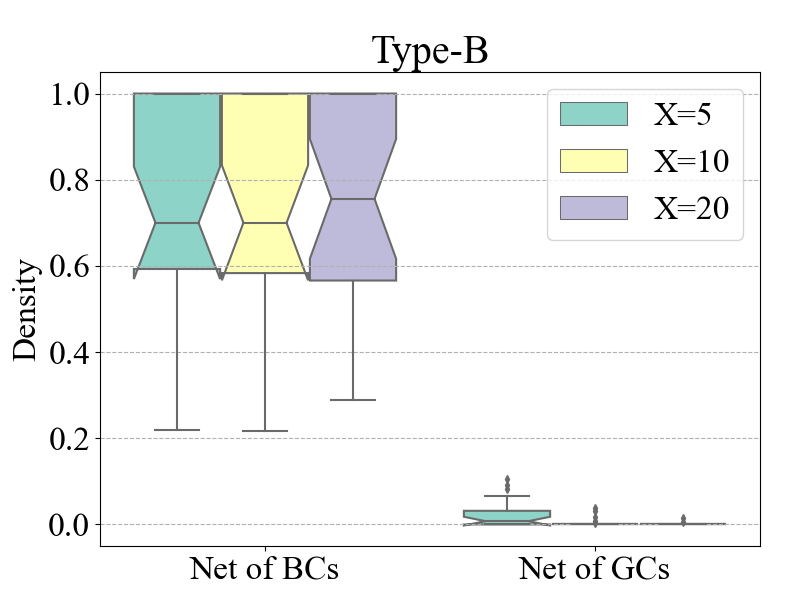}%
		\label{fig3densityB}}
	\caption{Density of sub-networks while $X$=5,10 and 20, (a) in the total network $G^{com}$, (b)  in the network $G^{Acom}$, (c)  in the network $G^{Bcom}$. }
	\label{fig3Density}
\end{figure*}

\begin{figure*}[t]
	\centering
	\subfigure[]{\includegraphics[width=0.325\textwidth]{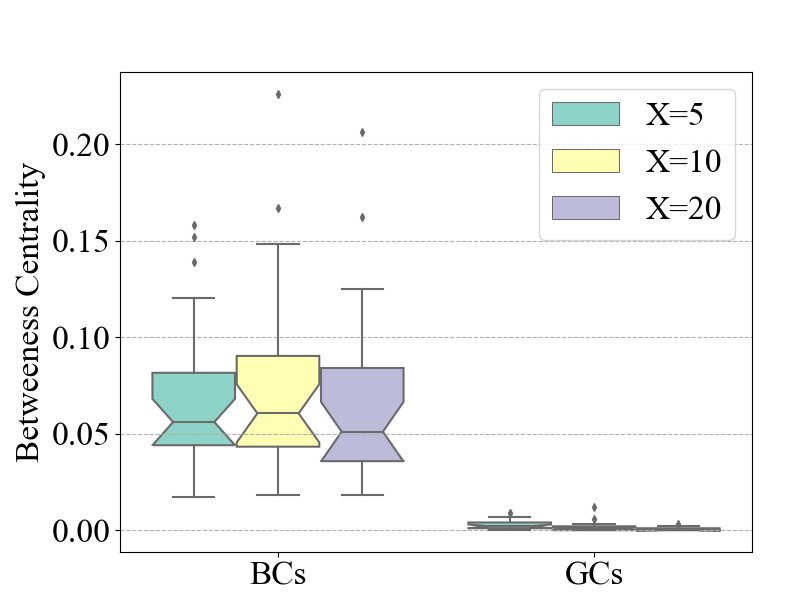}%
		\label{fig3bet}}
	\hfil
	\subfigure[]{\includegraphics[width=0.325\textwidth]{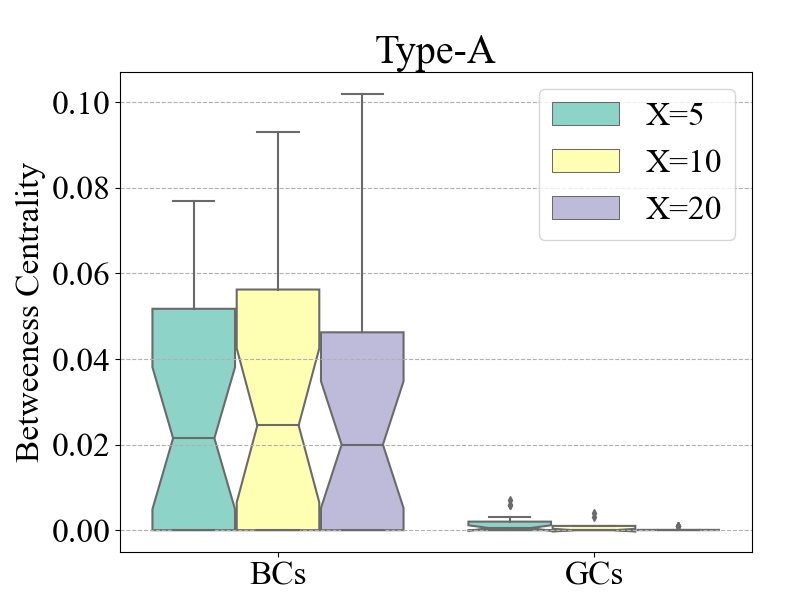}%
		\label{fig3betA}}
	\hfil
	\subfigure[]{\includegraphics[width=0.325\textwidth]{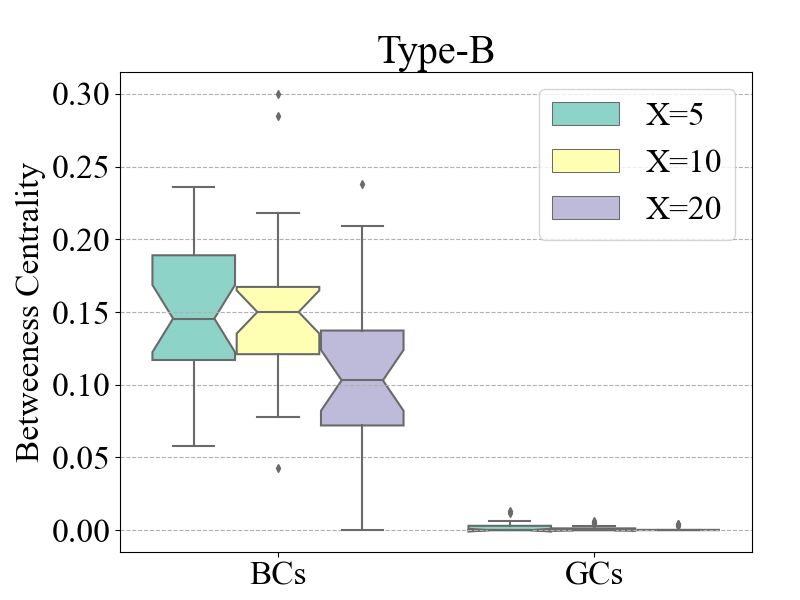}%
		\label{fig3betB}}
	
	\caption{Betweeness centrality of BCs and GCs in abstracted network while $X$=5,10 and 20, (a) in the total network $G^{com}$, (b)  in the network $G^{Acom}$, (c)  in the network $G^{Bcom}$.}
	\label{fig3Betweeness}
\end{figure*}

We present the graph of the abstracted network in some time frames while $X=10$ in Fig.~\ref{figCom}, where the red circles represent BCs and the green circles are GCs. The red, green and brown edges are BBEs, GGEs and BGEs, respectively. The size of the circles reflects the number of members in the community, and the thickness of the edges indicates the intensity of interaction between members of communities. 
It is interesting that the network, which is star-shaped, contains a dense cohesive core and a sparsely connected periphery. This observation is further verified in Fig.~\ref{figComEdge}, ~\ref{figDensity} and ~\ref{figBetweeness}.
It can be seen in Fig.~\ref{figComEdge}\subref{figcomnum5} \subref{figcomnum10} and \subref{figcomnum20} that the number of BCs is stable, while the number of GCs has a larger variation and follows a similar trend as the network size changes. The GCs out-numbers BCs in most time frames, but the former is far more sparse than the latter. Fig.~\ref{figComEdge}\subref{figedge5} \subref{figedge10} and \subref{figedge20} show the proportion of the weights of the three types of edges to the total weight of all edges. The edges are mostly between BCs or between BCs and GCs. Fig.~\ref{figDensity} presents the network density of the two sub-networks, containing either BCs and BBEs or GCs and GGEs. We can see that the BCs-formed sub-network has a high density, while the density of the sub-network formed by GCs is always low. In Fig.~\ref{figBetweeness}, the betweenness centrality of backbone members and general members are displayed respectively. By comparison, it can be found that backbone members have higher betweenness centrality, playing an important role as mediators, while GCs are isolated from each other and must be bridged by BCs. 
This also suggests that promoting communications between groups at the periphery may be effective to increase network connectivity.

A further look into the structure of the networks under type-A and type-B activities shows that the two networks also exhibit a clear core-periphery structure. With type-A activities, the number of connections between BCs and GCs (BGEs) significantly out-numbers that are among BCs (BBEs) or GCs (GGEs) (Fig.~\ref{fig2edge}\subref{fig2edgeA}), while with type-B activities, the number of BBEs out-numbers that of BGEs (Fig.~\ref{fig2edge}\subref{fig2edgeB}). This indicates that while BMs tend to be active in both types of activities, they may behave very differently. In type-A, i.e., rewarding activities, BMs are more likely to connect with GMs, but in type-B activities, BMs tend to connect with other BMs. This on the first hand validates the trunk-like function that the BMs play. On the other hand, this suggests that even though rewarding activities are generally more popular than non-rewarding ones and participation is much higher, the backbone members are still important convening points for the activities, and are thus very important to the development of the CSO. It can be seen in Fig.~\ref{fig3Density}\subref{fig3density}\subref{fig3densityA}\subref{fig3densityB} that in both types of activities, connections between GCs are very rare, again validating the leaf-like behavior of GCs and GMs. It also can be observed from Fig.~\ref{fig3Density}\subref{fig3densityB} that non-rewarding activities provide an important vehicle for backbone members to develop and consolidate very close relations, and should be regarded as an important tool in the entire CSO development toolset.

\section{Related Work}
Research in the domain of ``who are the most important nodes in the network" has considered various criteria to define influential participants. One stream of relevant research is node centrality \cite{das2018study}, such as degree centrality, closeness centrality and so on, which attempts to quantify the structural importance of actors in a network. Considering two metrics, the degree of the node and its position in the network topology, Kitsak $et\ al.$ \cite{kitsak2010identification} designed a k-shell decomposition method to divide the network nodes into different layers, i.e., to determine the importance of the nodes hierarchically at the group level. Subsequently, authors in \cite{garas2012k} introduced a generalized method
for calculating the k-shell structure of weighted networks. Our work goes a step further to improve the existing weighted k-shell method by taking the temporal nature of the network into account.

Taking influential nodes as research objects, researchers have drawn many interesting conclusions. Kerlund \cite{2020The} shows that influential users have a narrow focus in terms of the content they post and how they profile themselves and tend to produce more original content than other users. 
Zhao $et\ al.$ \cite{zhao2021propagation} find that for the entertainment news, the influential spreaders may appear at the later stage of spreading. Borgatti $et\ al.$ present an intuitive description of the core-periphery structure in \cite{rombach2014core}, that is, the network contains a dense, cohesive core and a sparse, unconnected periphery, and then quantified the core-periphery structure using the quadratic assignment procedure. Authors of \cite {shafiq2013identifying} performed a detailed analysis on the key topological properties of the friendship graph for three different user categories of Leaders, Followers and Neutrals and yielded interesting insights. Wang $et\ al.$ \cite{wang2020public} divided the nodes in the network into two categories: central nodes and ordinary nodes, further they found that the information dissemination mode can be summarized into three specific patterns. By decomposing the complex network structure into different parts and analyzing them systematically, it is possible to grasp the development pattern of the network in more detail.

Social Network Analysis (SNA)\cite{marin2011social}, focusing on understanding the nature and consequences of relations between individuals or groups, has been widely used to study social networking platforms.
Newman proposed that social networks can be naturally divided into communities or modules in \cite{newman2006modularity}. Blondel $et\ al.$ \cite{blondel2008fast} proposed a method known as the Louvain algorithm for community detection, with the core idea of optimizing the quality function known as modularity. Frequent changes in the activity and communication patterns of network members result in the associated social and communication network being subject to constant evolution. For a deeper understanding of network development, Palla $et\ at.$ quantified network evolution from the perspective of social group evolution in \cite{palla2007quantifying}. The evolution events are specifically classified into seven categories by group evolution discovery based on the changes of members in the communities in \cite{brodka2013ged}. Based on this, \cite{zhou2020formation, yu2021community} introduced an improved GED algorithm, describing network evolution in the context of CSOs. These works, from the perspective of community evolution in networks, have inspired the analysis of dynamic network structures on a vertical timeline. 

\section{Conclusions}
We propose a new framework called Twotier for analyzing the network structure and apply it to probe a non-profit sports organization. Firstly we establish a time-evolving network based on the team-wise relationships of participants. Taking the degree and topological position of nodes as well as the temporal nature of the CSO into account, we extend the weighted k-shell decomposition to determine the influence of the nodes and next classify the participants into two categories: backbone members and general members. Further the network of communities is abstracted by hiding the connections within communities. We not only analyze the development of the organization from the perspective of community evolution on the vertical timeline, but also pay attention to the connections between different groups in the horizontal time frames. In addition we have discussed the effect of the external stimulus on both groups.
Our findings are summarized as follows.

The backbone members of the CSO are not only characterized by their high degree and closeness centrality, but also by the fact that the average number of participating activities and active time frames of them are much higher than the general members. On average, the participation of backbone members in non-rewarding activities is higher than in rewarding activities, being the opposite of the general members.

Through Tier-one analysis, we reveal that the two sub-networks, containing either backbone or general members, both have a clear community structure. The groups of backbone members play the role of the trunk, while the general members renew frequently and act like leaves. 
Through Tier-two analysis, we identify that there is a core-periphery structure in the organization. Backbone members serve as a critical link between different groups within an organization, and organizational leaders should pay special attention to their role in managing the organization effectively. However, we also note the potential negative impact of few ties between communities of general members on membership stability and organizational development. Therefore, it is necessary to implement measures to strengthen interaction between these groups and break down isolation.

More importantly, we observe that external stimulus affects the organization in different ways. Even though rewarding activities are generally more popular than non-rewarding ones and participation is much higher, the backbone members are still important convening points for the activities, and are thus very important to the development of the CSO. The non-rewarding activities provide an important vehicle for backbone members to develop and consolidate very close relations and should be regarded as an important tool in the entire CSO development toolset. These insights can help practitioners develop tailored approaches for different groups within their organization to ensure better outcomes.

\bibliographystyle{IEEEtran}
\bibliography{IEEEexample}

\begin{thebibliography}{10}
\providecommand{\url}[1]{#1}
\csname url@samestyle\endcsname
\providecommand{\newblock}{\relax}
\providecommand{\bibinfo}[2]{#2}
\providecommand{\BIBentrySTDinterwordspacing}{\spaceskip=0pt\relax}
\providecommand{\BIBentryALTinterwordstretchfactor}{4}
\providecommand{\BIBentryALTinterwordspacing}{\spaceskip=\fontdimen2\font plus
\BIBentryALTinterwordstretchfactor\fontdimen3\font minus
  \fontdimen4\font\relax}
\providecommand{\BIBforeignlanguage}[2]{{%
\expandafter\ifx\csname l@#1\endcsname\relax
\typeout{** WARNING: IEEEtran.bst: No hyphenation pattern has been}%
\typeout{** loaded for the language `#1'. Using the pattern for}%
\typeout{** the default language instead.}%
\else
\language=\csname l@#1\endcsname
\fi
#2}}
\providecommand{\BIBdecl}{\relax}
\BIBdecl

\bibitem{misener2014support}
K.~Misener and A.~Doherty, ``In support of sport: Examining the relationship
  between community sport organizations and sponsors,'' \emph{Sport Management
  Review}, vol.~17, no.~4, pp. 493--506, 2014.

\bibitem{van2020community}
K.~Van~der Veken, E.~Lauwerier, and S.~J. Willems, ``How community sport
  programs may improve the health of vulnerable population groups: a program
  theory,'' \emph{International journal for equity in health}, vol.~19, no.~1,
  pp. 1--12, 2020.

\bibitem{westberg2022promoting}
K.~Westberg, C.~Stavros, L.~Parker, A.~Powell, D.~M. Martin, A.~Worsley,
  M.~Reid, and D.~Fouvy, ``Promoting healthy eating in the community sport
  setting: A scoping review,'' \emph{Health Promotion International}, vol.~37,
  no.~1, p. daab030, 2022.

\bibitem{doherty2019organizational}
A.~Doherty and G.~Cuskelly, ``Organizational capacity and performance of
  community sport clubs,'' \emph{Journal of Sport Management}, vol.~34, no.~3,
  pp. 240--259, 2019.

\bibitem{zhu2017new}
J.~Zhu, Y.~Liu, and X.~Yin, ``A new structure-hole-based algorithm for
  influence maximization in large online social networks,'' \emph{IEEE Access},
  vol.~5, pp. 23\,405--23\,412, 2017.

\bibitem{2020The}
M.~Kerlund, ``The importance of influential users in (re)producing swedish
  far-right discourse on twitter:,'' \emph{European Journal of Communication},
  vol.~35, no.~6, pp. 613--628, 2020.

\bibitem{zhao2021propagation}
Z.~Zhao, ``Propagation structure feature of entertainment news in the weibo
  online social network,'' \emph{EPL (Europhysics Letters)}, vol. 135, no.~1,
  p. 16002, 2021.

\bibitem{zhao2014finding}
K.~Zhao, J.~Yen, G.~Greer, B.~Qiu, P.~Mitra, and K.~Portier, ``Finding
  influential users of online health communities: a new metric based on
  sentiment influence,'' \emph{Journal of the American Medical Informatics
  Association}, vol.~21, no.~e2, pp. e212--e218, 2014.

\bibitem{lu2016vital}
L.~L{\"u}, D.~Chen, X.-L. Ren, Q.-M. Zhang, Y.-C. Zhang, and T.~Zhou, ``Vital
  nodes identification in complex networks,'' \emph{Physics Reports}, vol. 650,
  pp. 1--63, 2016.

\bibitem{de2019general}
P.~De~Meo, M.~Levene, F.~Messina, and A.~Provetti, ``A general centrality
  framework-based on node navigability,'' \emph{IEEE Transactions on Knowledge
  and Data Engineering}, vol.~32, no.~11, pp. 2088--2100, 2019.

\bibitem{yu2022epidemic}
P.-D. Yu, C.~W. Tan, and H.-L. Fu, ``Epidemic source detection in contact
  tracing networks: Epidemic centrality in graphs and message-passing
  algorithms,'' \emph{IEEE Journal of Selected Topics in Signal Processing},
  vol.~16, no.~2, pp. 234--249, 2022.

\bibitem{santoro2022onbra}
D.~Santoro and I.~Sarpe, ``Onbra: Rigorous estimation of the temporal
  betweenness centrality in temporal networks,'' in \emph{Proceedings of the
  ACM Web Conference 2022}, 2022, pp. 1579--1588.

\bibitem{das2018study}
K.~Das, S.~Samanta, and M.~Pal, ``Study on centrality measures in social
  networks: a survey,'' \emph{Social network analysis and mining}, vol.~8,
  no.~1, pp. 1--11, 2018.

\bibitem{rombach2014core}
M.~P. Rombach, M.~A. Porter, J.~H. Fowler, and P.~J. Mucha, ``Core-periphery
  structure in networks,'' \emph{SIAM Journal on Applied mathematics}, vol.~74,
  no.~1, pp. 167--190, 2014.

\bibitem{kitsak2010identification}
M.~Kitsak, L.~K. Gallos, S.~Havlin, F.~Liljeros, L.~Muchnik, H.~E. Stanley, and
  H.~A. Makse, ``Identification of influential spreaders in complex networks,''
  \emph{Nature physics}, vol.~6, no.~11, pp. 888--893, 2010.

\bibitem{garas2012k}
A.~Garas, F.~Schweitzer, and S.~Havlin, ``A k-shell decomposition method for
  weighted networks,'' \emph{New Journal of Physics}, vol.~14, no.~8, p.
  083030, 2012.

\bibitem{zhou2020formation}
Q.~Zhou, J.~Yu, and W.~Sun, ``Formation of a community: in the case of a
  particular non-profit sports organization,'' in \emph{2020 International
  Conference on Computing, Networking and Communications (ICNC)}.\hskip 1em
  plus 0.5em minus 0.4em\relax IEEE, 2020, pp. 844--848.

\bibitem{yu2021community}
J.~Yu, M.~Ding, Q.~Wang, W.~Sun, and W.~Hu, ``Community sports organization
  development from a social network evolution perspective--structures, stages,
  and stimulus,'' \emph{IEEE Transactions on Computational Social Systems},
  2021.

\bibitem{shafiq2013identifying}
M.~Z. Shafiq, M.~U. Ilyas, A.~X. Liu, and H.~Radha, ``Identifying leaders and
  followers in online social networks,'' \emph{IEEE Journal on Selected Areas
  in Communications}, vol.~31, no.~9, pp. 618--628, 2013.

\bibitem{ott2018strategic}
M.~Q. Ott, J.~M. Light, M.~A. Clark, and N.~P. Barnett, ``Strategic players for
  identifying optimal social network intervention subjects,'' \emph{Social
  networks}, vol.~55, pp. 97--103, 2018.

\bibitem{brodka2013ged}
P.~Br{\'o}dka, S.~Saganowski, and P.~Kazienko, ``{GED}: the method for group
  evolution discovery in social networks,'' \emph{Social Network Analysis and
  Mining}, vol.~3, no.~1, pp. 1--14, 2013.

\bibitem{blondel2008fast}
V.~D. Blondel, J.-L. Guillaume, R.~Lambiotte, and E.~Lefebvre, ``Fast unfolding
  of communities in large networks,'' \emph{Journal of statistical mechanics:
  theory and experiment}, vol. 2008, no.~10, p. P10008, 2008.

\bibitem{2009Mining}
H.~Kwak, Y.~Choi, Y.~H. Eom, H.~Jeong, and S.~B. Moon, ``Mining communities in
  networks: A solution for consistency and its evaluation,'' in
  \emph{Proceedings of the 9th ACM SIGCOMM Conference on Internet Measurement
  2009, Chicago, Illinois, USA, November 4-6, 2009}, 2009.

\bibitem{wang2020public}
X.~Wang, Y.~Xing, Y.~Wei, Q.~Zheng, and G.~Xing, ``Public opinion information
  dissemination in mobile social networks--taking sina weibo as an example,''
  \emph{Information Discovery and Delivery}, 2020.

\bibitem{marin2011social}
A.~Marin and B.~Wellman, ``Social network analysis: An introduction,''
  \emph{The SAGE handbook of social network analysis}, vol.~11, p.~25, 2011.

\bibitem{newman2006modularity}
M.~E. Newman, ``Modularity and community structure in networks,''
  \emph{Proceedings of the national academy of sciences}, vol. 103, no.~23, pp.
  8577--8582, 2006.

\bibitem{palla2007quantifying}
G.~Palla, A.-L. Barab{\'a}si, and T.~Vicsek, ``Quantifying social group
  evolution,'' \emph{Nature}, vol. 446, no. 7136, pp. 664--667, 2007.

\end{thebibliography}

\end{document}